\documentclass[aps,pre,twocolumn,preprintnumbers,showpacs,floatfix,longbibliography]{revtex4-1}\usepackage{graphicx}
\usepackage{subfig}
\usepackage{amsmath}
\usepackage{amssymb}
\usepackage{gensymb}
\usepackage{float}
\usepackage{color}
\usepackage{multirow}
\usepackage{soul}

\begin{document}
\title{Properties of Interaction Networks in Compressed Two and Three Dimensional Particulate Systems}

\author{L. Kovalcinova}\altaffiliation[Current address:]{ Google, Inc., New York, NY}
\affiliation{Department of Mathematical
  Sciences, New Jersey Institute of Technology, Newark, NJ}
  
  \author{A. Taranto}
\affiliation{Department of Mathematical
  Sciences, New Jersey Institute of Technology, Newark, NJ}
  
  \author{L. Kondic} \email[Corresponding author:]{ kondic@njit.edu}
\affiliation{Department of Mathematicalhttps://www.overleaf.com/project/5c28d1df4a493c1181d14c6c
  Sciences, New Jersey Institute of Technology, Newark, NJ}

\date{\today}

\begin{abstract}
We consider two (2D) and three (3D) dimensional granular systems 
exposed to compression, and ask what is the influence of the number of physical dimensions on the properties of the interaction networks that 
spontaneously form as these systems evolve.  The study is carried out 
based on discrete element simulations of frictional disks in 2D and 
spheres in 3D. 
Within the constraints of the considered protocols involving compression, system 
properties (involving the presence of bounding walls) and finite system sizes considered,
the main finding is that both the number of physical dimensions and the type of particle-particle interaction influence significantly the properties of interaction networks.  These networks play an important role in bridging the microscale (particle size) and macroscale (system size), thus both aspects (the interaction model and dimensionality) 
are carefully considered. Our work uses a combination of tools and techniques, including percolation  study, statistical analysis, as well as algebraic topology-based techniques.  In many instances we find that different techniques and measures provide complementary information that, when 
combined, allow for gaining  better insight into the properties of interaction networks. 
\end{abstract}
\maketitle

\section{Introduction}

Particulate systems are very common in nature and take different form ranging from  
dry or wet granular matter to suspensions. Still, there are many unresolved fundamental questions. 
  In particular, the connection between microscale behavior (on the scale of constitutive
  granular particles) and macroscale response of granular systems that are subject to external
  load, such as compression or shear, is still an open question.
  
A significant body of work, that has been mostly developed during the last decade, suggests that 
connecting  the micro and macroscale involves interaction networks that spontaneously develop
in particulate-based systems.  These works involve topology-based 
studies~\cite{arevalo_bif_chaos_09, arevalo_pre10, arevalo_pre13, azema_pre_12, ardanza_pre14,epl12, pre13,kramar_2014,kramar_pre_14,Pugnaloni_2016,kondic_2016,pg17,PRE_2018,pre18_impact}, 
cluster analysis~\cite{zhang_10,tordesillas_pre10,walker_pre12},
community detection schemes~\cite{peters05,daniels_pre12}, force network 
ensembles~\cite{tighe_sm10,tighe_jsm11,sarkar_prl13}, statistical analysis~\cite{radjai_96b,radjai97b,radjai99}
percolation-based
approaches~\cite{ostojic06, kovalcinova_scaling, ostojic_pre07,stauffer}, and discussion of percolation and jamming
transitions~\cite{kovalcinova_percolation, ohern01, ohern_pre03, ohern_pre_12, Pastor-Satorras_12, stauffer}.

\begin{figure}[ht!]
  \centering
  \subfloat[]{\includegraphics[width=2.5in]{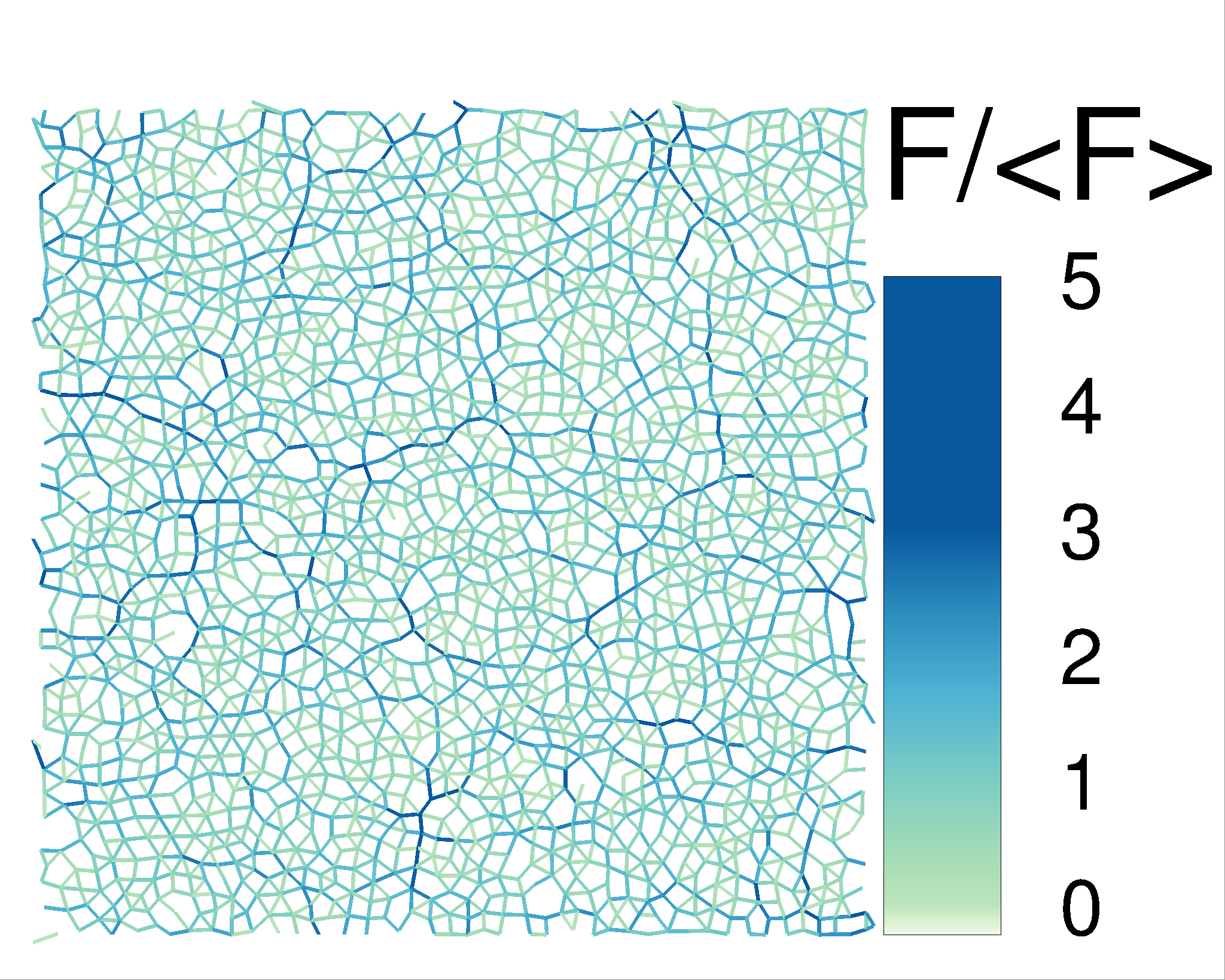}}\\
  \subfloat[]{\includegraphics[width=2.5in]{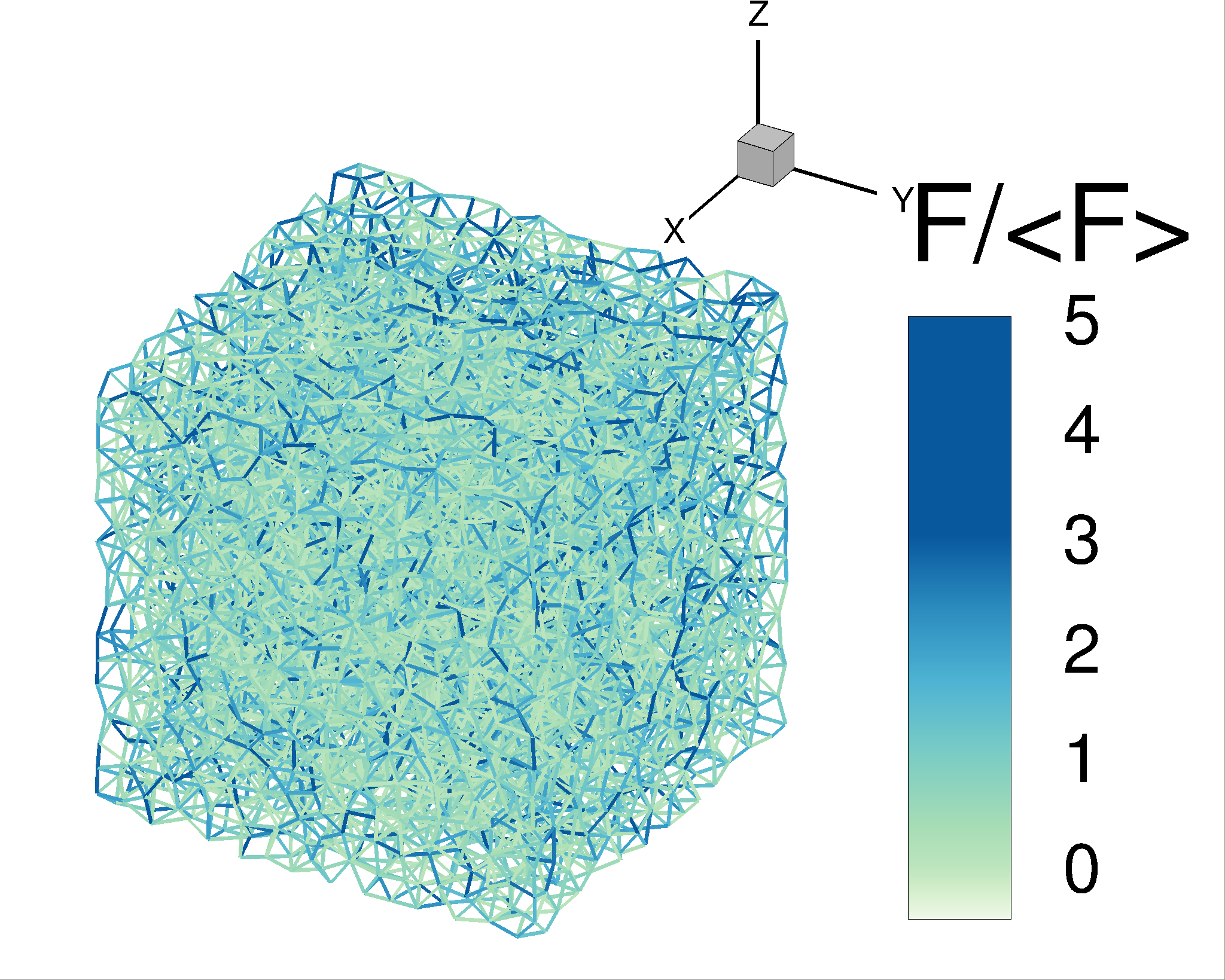}}\\
  \caption{Examples of interaction networks for the (a) $2$D and (b) $3$D granular systems.   
$\langle F \rangle$  is the average force.    
Animations of the 
  compression process  are available as 
  Supplementary Material~\cite{supp_mat}.
}\label{fig:FN}
\end{figure}

Most of the past analysis of interaction networks obtained computationally is carried out in two spatial dimensions ($2$D) due to reduced computational complexity and simplified interpretation of the results. 
A significant amount of experimental work focusing on interaction networks has been carried out in $2$D as well, mostly 
using photoelastic particles, see, e.g.~\cite{majmudar05a,clark_prl12}.  
However, a majority of applications focus on three spatial dimensions ($3$D).  In $3$D experiments, it is difficult to extract the information about particle contacts and in particular interaction strength.   While there have been recent experimental works in $3$D using, e.g., hydrogel~\cite{brodu_natcom_15} rubber~\cite{Saadatfar_2012} particles,
or oil droplets~\cite{brujic_2003}, the progress has been rather limited.
Currently, experimental techniques that could be used to track the force
 networks for stiff and frictional 3D particles, are just being developed~\cite{hurley_prl16}, and this field of investigation has to rely 
 mostly on simulation results. 
 In addition to being computationally costly,  $3$D simulations are complicated to
analyze, compared to 2D.    As an illustration, Fig.~\ref{fig:FN} shows two examples from the simulations that we will consider in the present paper.
Relatively recent approaches that focus on inter-particle forces include measurements of force probability density function~\cite{silbert02a}, as well as various network-based approaches (see~\cite{Papadopoulos_18} for a review). However, it is not obvious how to 
correlate and compare $3$D results to the ones obtained in much more commonly considered $2$D geometry.  
This is the main goal of the present work.  

The number of physical dimensions has complex influence on the properties of interaction networks in 
particulate systems.  First, the interaction networks are built on top of  contact networks between 
the particles, and contact networks are clearly influenced by the number of physical dimensions, 
since the geometry of particle contacts differs in $2$D and $3$D.   Second, the interaction between the 
particles is influenced by the particle shape - commonly disks are considered in $2$D and spheres 
in $3$D (of course, more complex $2$D and $3$D shapes could be considered, but we do not discuss
these in the present work).   Assuming validity of the standard approach based on Hertz model, one finds 
that the (normal) interaction force, $F$, between particles should scale as $F \propto x^\delta$, where the compression distance, $x$, is defined in soft particle simulations as the difference between the sum of particle radii and the distance
of their centers, and $\delta = 1.0,~1.5$ for 2D and 3D, respectively.    Thus, there are two obvious aspects 
of the influence of the number of physical dimensions
on particle interactions.  It is desirable to separate these two aspects so that one could distinguish
between the influence of the number of dimensions on contact geometry, and on the force interaction law.  In addition,  experiments
carried out in $2$D in some cases find that particle interaction law deviates from Hertzian theory, and that 
$\delta \approx 1.4 - 1.5$~\cite{PRE_2018}.   Therefore, motivated both by our desire to separate the aspects influencing particle interactions, and by the listed experiments, 
 we consider $2$D systems with both $\delta = 1$ and $\delta = 1.5$, in addition to considering $3$D systems
 with $\delta = 1.5$.   
 
\begin{figure}[th!]
  \centering
   \subfloat[]{\includegraphics[width=2.5in]{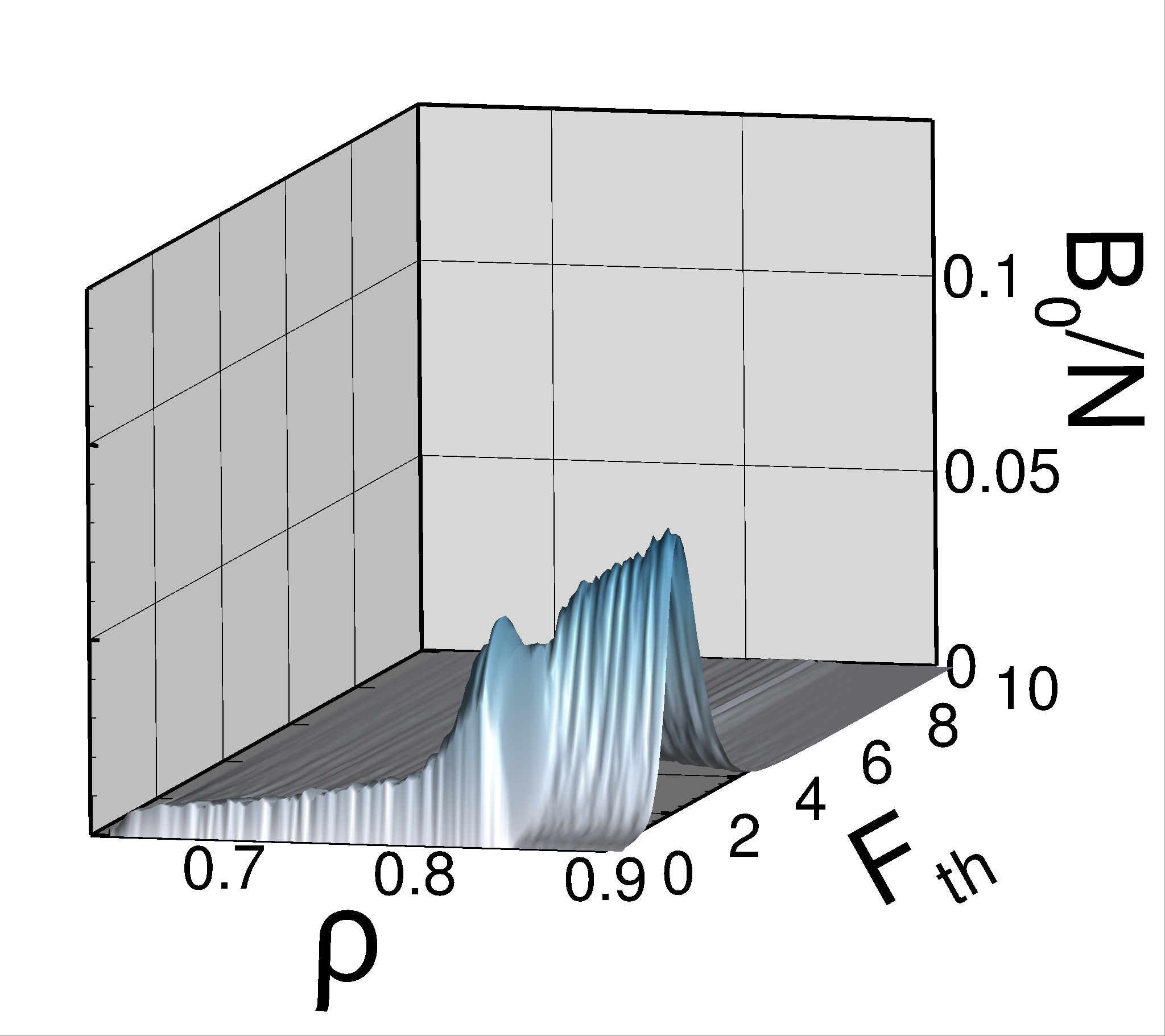}}\\
  \subfloat[]{\includegraphics[width=2.5in]{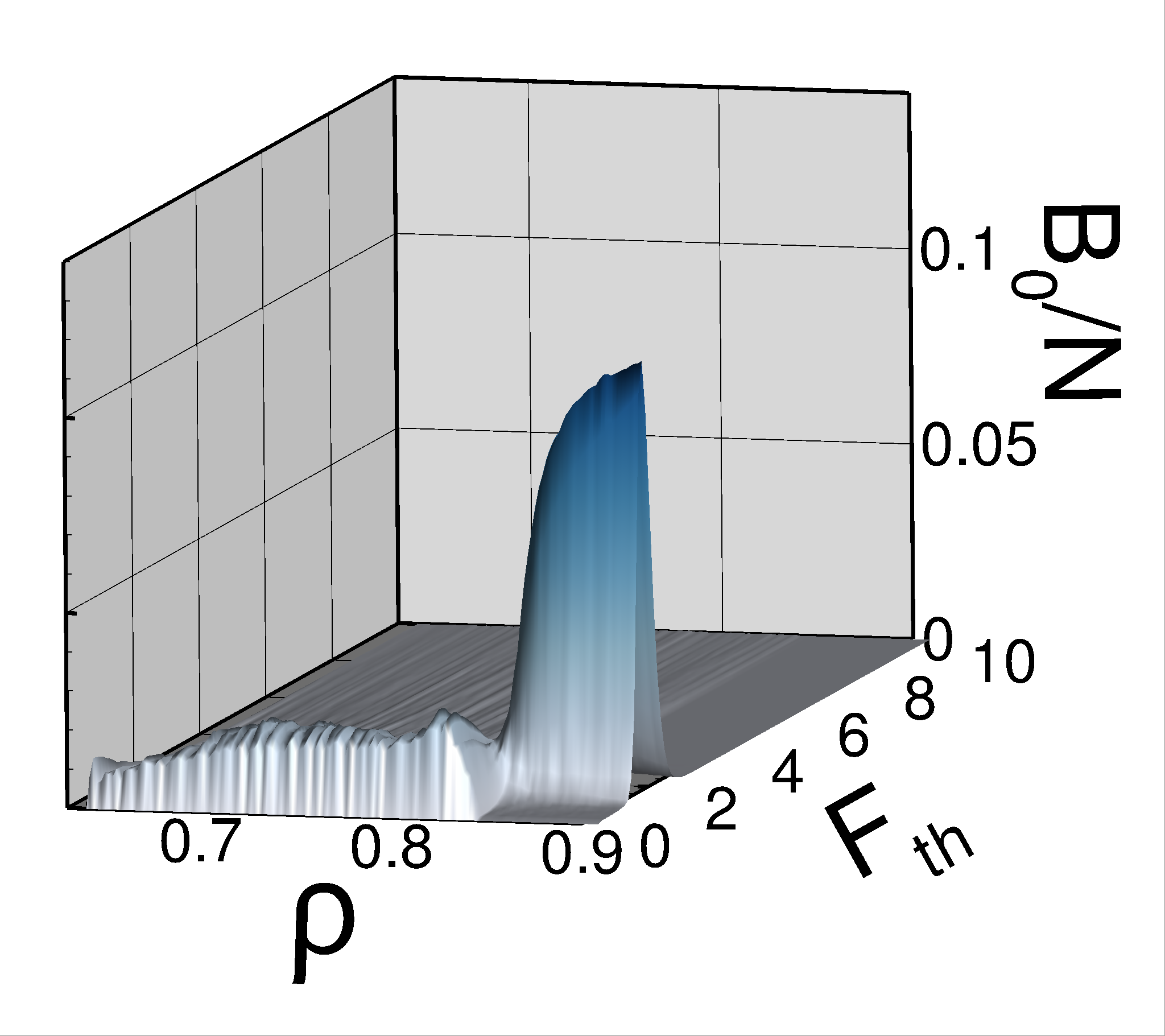}}\\
  \subfloat[]{\includegraphics[width=2.5in]{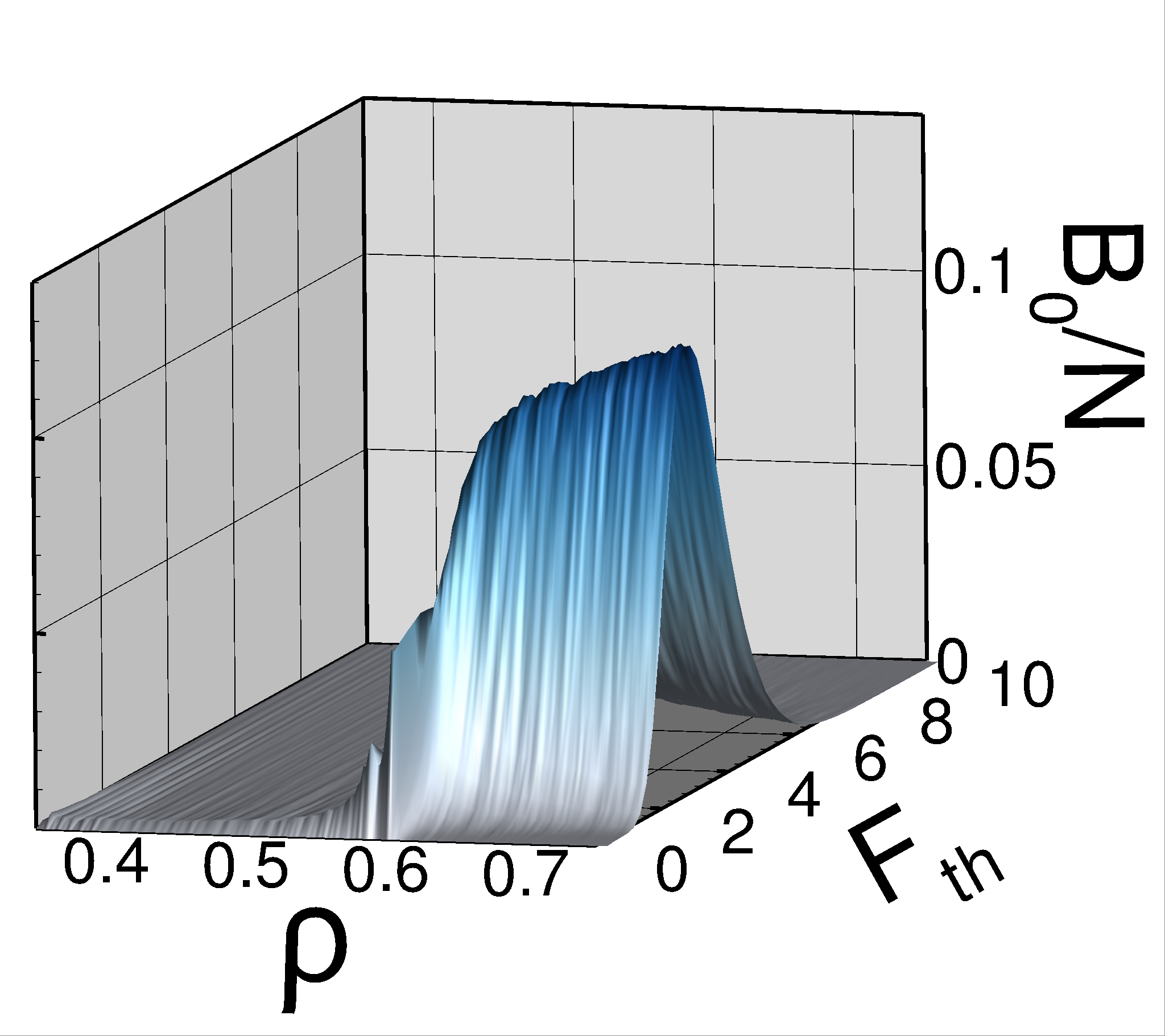}}
  \caption{The number of clusters in the interaction network, $B_0$,
  normalized by the total number of particles, $N$, for different values of packing fraction,
  $\rho$, and force threshold, $F_{\rm th}$ for (a) $2$D non-linear force model,  (b) 2D linear force model and (c) $3$D system.  
  The animations of different views of the panels are available as 
  Supplementary Material~\cite{supp_mat}.  
  }\label{fig:B0}
\end{figure}

In this paper, we focus on granular systems, however we expect that the results will be of interest to other 
particulate-based systems, such as suspensions, emulsions, or foams, and even to the systems characterized by more 
complex interaction between interacting particles, such as gels.  We focus on the compression protocol, so that we 
could discuss evolution of interaction networks as the considered systems evolve through jamming transition.    

To introduce the results that will be discussed in the rest of this paper, Fig.~\ref{fig:B0} shows the 
simplest topological measure, zeroth Betti number, $B_0$, that counts the number of connected components (clusters), 
as a function of imposed force threshold (panels (c - e))  (only the particle contacts characterized by the force larger
than the specified threshold are considered).  The evolution from small to large volume fractions is driven 
by slowly converging walls, as discussed in more detail in Sec.~\ref{sec:protocol}.  
To put the results in perspective, we note that choosing 
a threshold, $F_{\rm th}$, that corresponds to the average force between the particles, would 
separate the considered networks into `strong' (the contacts involving forces larger than the average), and 
`weak' (the remaining contacts)~\cite{radjai97}.  The approach considered
in the present work is more general, since we vary the considered force thresholds continuously and are therefore 
in the position to obtain more complete picture describing the networks.  We will discuss $B_0$ introduced above, as well as 
$B_1$, that measures the number of loops in $2$D, or tunnels in $3$D (sometimes also called `cycles').   
In $3$D we will also discuss $B_2$, measuring the number of enclosed voids, but to much lesser extend.    
In the second part of the paper we present the results obtained by implementing more
complex measures based on persistent homology.  We note that the software for computing Betti numbers as well
as persistence measures discussed later in the paper is widely available in the public 
domain~\cite{perseus,javaPlex,phat,gudhi}.

Before closing the introduction, we should also point out the aspects of the considered problem 
that are addressed only marginally in the present work.  First, the main part of results that are reported in 
the manuscript are obtained using specified domain size for the 2D and 3D systems considered.  
We carried out limited set of additional simulations and corresponding data analysis to verify that the 
results regarding interaction networks remain valid for other system sizes, but have not 
carried out extensive analysis, leaving such endeavor for the future work.  Our focus in the 
present work is on the qualitative properties of interaction networks in the systems that 
differ in the number of physical dimensions and the interaction model.  We point out to the 
reader that there is a a number of works where the influence of system size has been discussed; for the
discussion in the context of interaction networks we refer the reader to~\cite{ostojic,kovalcinova_scaling} and 
references therein; carrying out similar type of analysis with focus on the detailed 
properties of interaction networks remains to be done.   Second, most of the results
that we report in the manuscript are empirical, since we are not aware of a well-defined
theory or at least a model that could be used for comparison in the context of connectivity
and structure of the interaction networks.  This being said, we attempt
where possible to provide at least qualitative arguments rationalizing the physical 
reasons leading to the observed properties 
of the interaction networks. 

The rest of this paper is organized as follows.   The force model describing the interaction between particles is 
discussed in Sec.~\ref{sec:force}, and the protocol of the simulations in Sec.~\ref{sec:protocol}.   Section~\ref{sec:results1} 
gives results based on consideration of percolation, jamming, and force distributions. A complementary set of 
results obtained using topology-based methods is discussed in Sec.~\ref{sec:topology}.   An overview and conclusions 
are given in Sec.~\ref{sec:conclusions}.

\section{Force Model}
\label{sec:force}
We perform discrete element simulations using a set of circular disks confined to a square domain ($2$D) or spheres in a
 cubic domain ($3$D). The number of particles in $2$D is $\approx 2000$ and $\approx 4000$ in $3$D.  The walls are flat 
 and are given material properties of a plexiglass (the choice of the wall material is motivated by recent experiments~\cite{brodu_natcom_15}). 
 The domain length is initially set to $50\times 50$ (in $2$D) and $20\times 20\times 20$ (in $3$D) average particle diameters. 
 To avoid crystallization, the system particles are
 chosen as bidisperse in all systems and the ratio of the diameters of the small and large particles is $1:1.4$.
The ratio of the number of small to large particles is $2:1$.  

Particles are soft and interact via normal and tangential forces that include static friction and viscous damping.
The force model used in the simulations is either linear or non-linear. The linear force model is used only in $2$D and 
arises from the derivation of the Hertz law. The non-linear force is considered for both $2$D and $3$D. The analysis of 
the results for the non-linear force model in $2$D is motivated by our recent study~\cite{PRE_2018} that 
showed that using non-linear force model in $2$D is essential for achieving a quantitative match between simulations and experiments.

The general form of the normal force between particles $i$ and $j$ is (for more details, see~\cite{kondic_99})
\begin{eqnarray}
 {\bf F}_{\rm i,j}^{\rm n}& =&k_{\rm n}x^{\delta} {\bf n} - \gamma_{\rm n} x^{\nu} \bar m {\bf v}_{\rm i,j}^{\rm n} \nonumber\\
 &&\rm{if} \hspace{0.2cm}  ||k_{\rm n}x^{\delta} {\bf n}|| \geq ||\gamma_{\rm n} x^{\nu} \bar m {\bf v}_{\rm i,j}^{\rm n}||\nonumber\\
 &=& {\bf 0} \hspace{0.2cm}  \rm{otherwise}  \label{eq:non_lin_fm}\\
 r_{\rm i,j} &=& |{\bf r}_{\rm i,j}|, \ \ \ {\bf r}_{\rm i,j} = {\bf r}_{\rm i} - {\bf r}_{\rm j}, \ \ \  {\bf n} = {\bf r}_{\rm i,j}/r_{\rm i,j}\nonumber \\
 k_{\rm n} &=& {2Y\over 3(1-\sigma^2)}d_{\rm ave}^{1-\beta} \nonumber
\end{eqnarray}
Exponents $\delta = 1$ and $\nu = 0$ for linear and $\delta = 1.5$ and $\nu=0.5$ for non-linear force model 
with  $1+\beta=\delta$; ${\bf v}_{\rm i,j}^{\rm n}$ is set to the relative normal velocity and $Y$ and $\sigma$ are Young's 
modulus and Poisson ratio, respectively. The amount of compression is $x = d_{\rm i,j}-r_{\rm i,j}$, where $d_{\rm i,j} = {(d_{\rm i} + d_{\rm j})/2}$,
$d_{\rm i}$ and $d_{\rm j}$ are the diameters of the particles $i$ and $j$.  Note that {for simplification we assume that} $d_{\rm ave}$
in the expression for the force constant, $k_{\rm n}$, is the overall average particle diameter of all particles.  Also, $r_{\rm i,j} = |{\bf r}_{\rm i} - {\bf r}_{\rm j}|$, 
where  ${\bf r}_{\rm i}, {\bf r}_{\rm j}$ are the vectors pointing from the centers of particles $i,j$ towards the point of contact.
 The value of Young's modulus $Y=23.45~\rm{KPa}$ and Poisson ratio $\sigma=0.5$  is set to those of
 soft hydrogel particles in~\cite{brodu_pre_15}.  Choosing the same values of material parameters for $2$D and $3$D simplifies
 the comparison between different systems that we consider; it should be also noted that the choice simply 
 specifies the time scale in the problem; a different choice of (for example) stiffer particles would simply change
 the relevant time scale without modifying the results in any significant manner \footnote{Minor influence of different 
 material properties of the particles could be due to particle - wall interaction, and, in the case of a nonlinear force model,
 due to weak dependence of the particle collision time on impact speed.}.

We implement the commonly used Cundall-Strack model for static friction~\cite{cundall79}, where a tangential 
spring is introduced between the particles for each new contact that forms at time $t=t_0$. Due to the relative 
motion of the particles, the spring length, ${\boldsymbol\xi}$, evolves as 
$\boldsymbol\xi=\int_{t_0}^{\rm t} {\bf v}_{\rm i,j}^{\rm t}~(t')~dt'$, where ${\bf v}_{\rm i,j}^{\rm t}= {\bf v}_{\rm i,j} - {\bf v}_{\rm i,j}^{\rm n}$ and ${\bf v}_{\rm i,j}$ is the relative velocity.
For long lasting contacts, $\boldsymbol\xi$ may not remain parallel to the current tangential direction defined by $\bf{t}={\bf v}_{\rm i,j}^{\rm t}/|{\bf v}_{\rm i,j}^{\rm t}|$
(see, e.g,.~\cite{brendel98}); we therefore define the corrected $\boldsymbol\xi{^\prime} = \boldsymbol\xi - \bf{n}(\bf{n} \cdot \boldsymbol\xi)$ and introduce the test force
\begin{equation}
{\bf F}^{\rm t*} =-k_{\rm t}x^{\beta}\boldsymbol\xi^\prime - \gamma_{\rm t} x^{\nu} \bar m {\bf v}_{\rm i,j}^{\rm t}
\end{equation}
where $k_{\rm t} = 6/7k_{\rm n}$ (close to the value used in~\cite{goldhirsch_nature05}) and $\gamma_{\rm t}$ is the coefficient of 
viscous damping in the tangential direction (with $\gamma_{\rm t} = {\gamma_{\rm n}}$). The value of the friction coefficient is set to 
$\mu = 0.5$. To ensure that the magnitude of the tangential force remains below the Coulomb threshold, we constrain the tangential force 
as follows
\begin{equation}
 {\bf F}^{\rm t} = \min(\mu |{\bf F}^{\rm n}|,|{\bf F}^{\rm t*}|){{\bf F}^{\rm t*}/|{\bf F}^{\rm t*}|}\, .
\end{equation}

The interaction between a particle and flat wall is given by the same expression as in the case of interaction between two particles except for assuming that the amount of compression $x = 0.5d_{\rm i} - r_{\rm i}$ where $d_{\rm i}$ is particle diameter and   $r_{\rm i} = |\boldsymbol{r}_{\rm i}|$ is the vector pointing from the particle center towards the point of contact with the flat wall.

All the results given in the rest of the paper are presented using the following length, time, and mass scales.  
The characteristic length scale is $d_{\rm ave}=1.735$ cm, and the average particle mass,
$\bar m = 3.0$ g~\cite{brodu_pre_15},
is the mass scale.  The binary particle collision time,  $\tau_{\rm c}$,
is the time scale set to~\cite{kondic_99}
\begin{equation}
  \tau_{\rm c} = \zeta(\beta) (1+0.5\beta)^{1\over 2+\beta} \Biggl(\bar m {3(1-\sigma^2)\over 2Yd_{\rm ave}^{(1-\beta)}}\Biggr)^{1\over 2+\beta}v_0^{-\beta\over2+\beta}
\end{equation}
where $v_0=10^{-2}$ cm/s is a characteristic magnitude of velocity in the
system (compression speed); prefactor $\zeta(\beta)$ is of the form~\cite{luding94d}
\begin{equation}
  \zeta(\beta) = {\sqrt{\pi}\Gamma(1/(2+\beta))\over(1+\beta/2)\Gamma((4+\beta)/(4+2\beta))}\, ,
\end{equation}
where $\Gamma$ denotes Gamma function. In particular, for the linear force model we
have $\zeta(0)=\pi$ and for the non-linear force model $\zeta(0.5)=2.94$.
The damping coefficient $\gamma_{\rm n}$
is obtained as reported in~\cite{kondic_99}. Note that the average particle mass and
diameter are the same for $2$D and $3$D systems; the thickness of the disk particles in 
$2$D is chosen to ensure that this is the case.

We integrate Newton's equations of motion for the translational and rotational
degrees of freedom using a $4$th order predictor-corrector method with the time step
$\Delta t =0.02\tau_{\rm c}$ and $\Delta t =0.005\tau_{\rm c}$ in $2$D and $3$D,
respectively.  Smaller value of time step in $3$D is needed to keep the minimum
distance between interacting particles sufficiently large for the whole duration of the 
simulations. The use of smaller time step in $3$D does not influence the results that
follow in any visible manner.

\section{Protocol and Methods}
\label{sec:protocol}

The initial condition is produced by placing the particles on a square (2D) or a cubic (3D)
grid and by assigning to each particle a random initial velocity; we verified that
the results presented here are not sensitive to the specific distribution from
which the velocities are sampled. To ensure a statistical significance of the results,
we perform simulations for $20$ different initial conditions for each considered system.  

The system is compressed by moving the walls
inward with the velocity $v_0$. Relaxation is interjected after each compression step
of $\Delta\rho=0.02$, where $\rho$ is the packing fraction, defined as the ratio of
the total volume of the particles and the domain volume. After each compression step, the kinetic
energy dissipates exponentially when the system is relaxed;  we continue relaxation
until the fluctuation of the kinetic energy drops to $0.1$ of the mean, that is computed over
$1.5\times 10^5$ time steps. We verified that a more strict relaxation condition does not change 
our conclusions nor does it change the measured quantities significantly; however if the 
threshold for the kinetic energy fluctuations is set to a value $<0.1$, the computational time
increases significantly and the differences in the results are on the level of
random fluctuations in each measured quantity. We find that using a 
protocol for relaxation based on the total magnitude of the kinetic energy does
not yield different results either.
We have also implemented different relaxation protocols based on tracking of the average contact 
number change and on cooling of the particles (annealing). The protocol  corresponding to the first method verifies 
that the contact number drops to a near-zero value for unjammed systems, and that for jammed systems
 the contact number  does not change over a large number of time steps.
We find that such protocol does not influence the results presented in this paper 
and it turns out to be more computationally expensive.  The annealing protocol 
implements the approach for cooling of the particles during relaxation as discussed recently~\cite{zhang_jchp_13}.
For the system considered in this work,  we find that jamming transition is influenced by cooling rate even for
the smallest cooling rates we could use with available computing resources,  and is therefore not suitable for 
our study.  

It is beyond the scope of the present work to discuss why implemented annealing protocol
has such an influence in the considered simulations; one possibility is that the differences between
implemented protocols vanish in the limit of large system sizes, however, as pointed out already, 
such an analysis is left for future work.

The reason for this detailed discussion of the implemented protocol is the presence of a relatively small, but non-vanishing, 
number of clusters for packing fractions below jamming in particular for $2$D simulations, see Fig.~\ref{fig:B0}, and 
Figs.~\ref{fig:avg_S} and~\ref{fig:B_0} later in the paper.  One obvious question
is whether these clusters would disappear if a different relaxation protocol were used.  Our finding is that 
these clusters remain present for any of the considered relaxation protocols. 

This observations is consistent with the earlier
works~\cite{Miller2004,Poeschel2005}  that found that the cooled unjammed systems always contain clusters
of connected particles, at least for the simulation setup and relaxation protocols considered in the present work.
One could also ask whether such clusters would be still present as the system size is increased, or 
whether the fraction of the particles participating in such clusters would decay as the system size increases.
Such questions should be considered in the future work.

{\it Interaction networks:} We consider the forces between particles and define the
interaction network in a granular system by its nodes (particle centers)  and weighted edges (inter-particle forces).    
The interaction networks are analyzed for different values of $\rho$ and for a range of (dimensionless) force thresholds, 
$F_{\rm th}\in[0.0,3.0]$, where only the forces that are larger than $F_{\rm th}$ are taken into account. We note 
that the forces are always rescaled by the average force in an interaction network, $\langle F\rangle$, which itself is 
$\rho$-dependent.   For simplicity, in this work we consider only the total force between the particles (the absolute
value of the vector sum of the normal and tangential forces at each contact).  
We show the results up to the value of $\rho$ such that the average force does not exceed a maximum, $F_{\rm max}$, 
defined as follows.  We find the average force in an interaction network, $\langle F\rangle$, for all $\rho$'s (averaged over 
$20$ realizations for $2$D linear and non-linear, and for $3$D systems),  and determine the value $F_{\rm max}'=\max \{\langle F\rangle\}$ 
for each considered system.   We define $F_{\rm max}=\min \{\langle F_{\rm max}'\rangle\}$ so that
for all considered systems, the results are given for  the same range of average force thresholds.
The  maximum packing fractions are found as $0.906,~0.890$ and $0.710$ for the 
non-linear $2$D system, linear 2$D$ system, and the $3$D system, respectively.  

{\it Data pre-processing:} During compression, especially for small values of $\rho$, the particles
 collide and often form two-particle clusters. The particles with only one or no contacts, are referred to as rattlers.
As in the previous experimental and numerical studies~\cite{PRE_2018, dapeng, Zhang_soft_2010},
we do not consider rattlers in our computations; these clusters are non included in the Betti number results discussed 
later in the paper.  

 We note that we have carried out additional simulations using different system 
sizes: $25\times 25$ and $75 \times 75$ (in $2$D) and $10\times 10\times 10$ and 
$30\times 30\times 30$ (in $3$D) average particle diameters. The finding of relevance 
for the present paper is that the results reported in the rest of manuscript remain
qualitatively the same for the considered domain sizes. We have also carried out simulations using 
different friction coefficient (in addition to the value $\mu = 0.5$ considered in all the reported simulations, 
we used $\mu =0.0$ and $\mu = 0.03$).   The only visible difference between systems with different friction values
was found in the specific value of $\rho_{\rm J}$, and in the precise values of the various measures describing the interaction networks.

\section{Force Distribution, Percolation and Jamming Transitions}
\label{sec:results1}

In this section we focus on the differences between $2$D and $3$D
systems that undergo jamming and percolation transitions (described below) in terms of the
distribution of forces and properties of the interaction networks.

During compression and relaxation, particles come into contact and
 clusters randomly form and break.  Since we include relaxation between compression steps, the compression protocol is quasi-static and we expect that the percolation and jamming transition occur at the same packing fraction~\cite{kovalcinova_percolation}. Note that both here and in~\cite{kovalcinova_percolation} a jamming transition is characterized by a rapid increase of average contact number. More specifically, jamming packing fraction,
$\rho_{\rm J}$, is defined here as the point at which $Z (\rho)$ curve has 
an inflection point. We refer the reader to recent works that discuss in more
detail current understanding of jamming in granular systems~\cite{sastry_nature16,luding_nature16}. 
Percolation transition is defined by $\rho = \rho_{\rm p}$  at which a percolating cluster connecting at 
least two opposite walls of the domain forms. We indeed observe the formation of a stable percolating cluster (that does not disappear even after 
arbitrarily long relaxation) for any (and only when) $\rho > \rho_{\rm J}$, confirming that $\rho_{\rm p}  = \rho_{\rm J}$.  
Finding that $\rho_{\rm p}  = \rho_{\rm J}$ also suggests
that the relaxation protocol that we use is essentially quasi-static, 
so that the results are not influenced in any meaningful manner by 
the relaxation protocol, see~\cite{kovalcinova_percolation} for further
discussion regarding this issue. 
For the future reference, we note that for the systems that we consider the jamming occurs at 
$\rho_{\rm J}=0.842,~0.834$ for the $2$D non-linear and
linear systems, respectively, and $\rho_{\rm J}=0.57$ for the $3$D system
{color{blue}
(note that these numbers may be influenced by the system size)}.

\begin{figure}[ht!]
  \centering
  \subfloat[]{\includegraphics[width=1.75in]{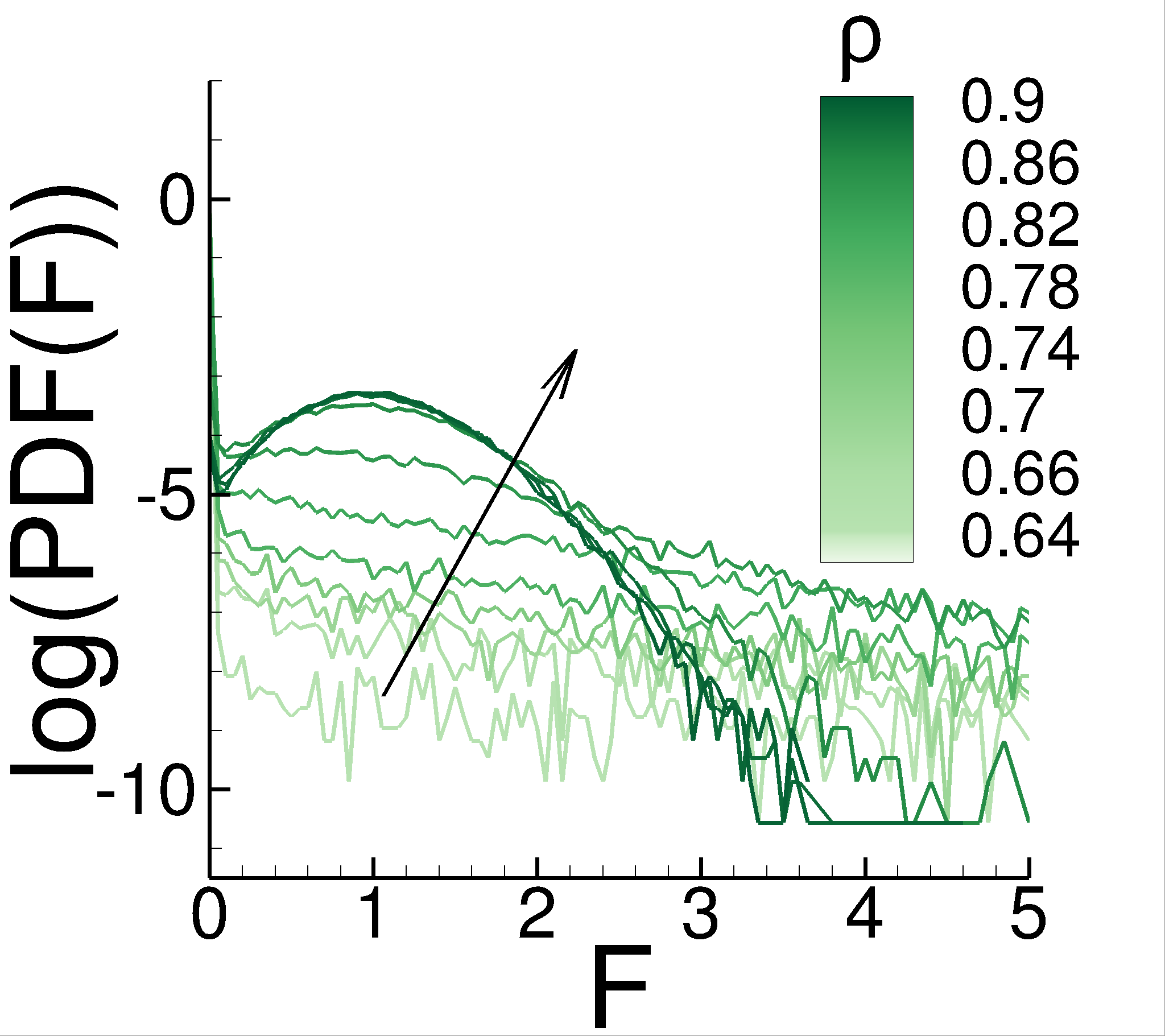}}
  \subfloat[]{\includegraphics[width=1.75in]{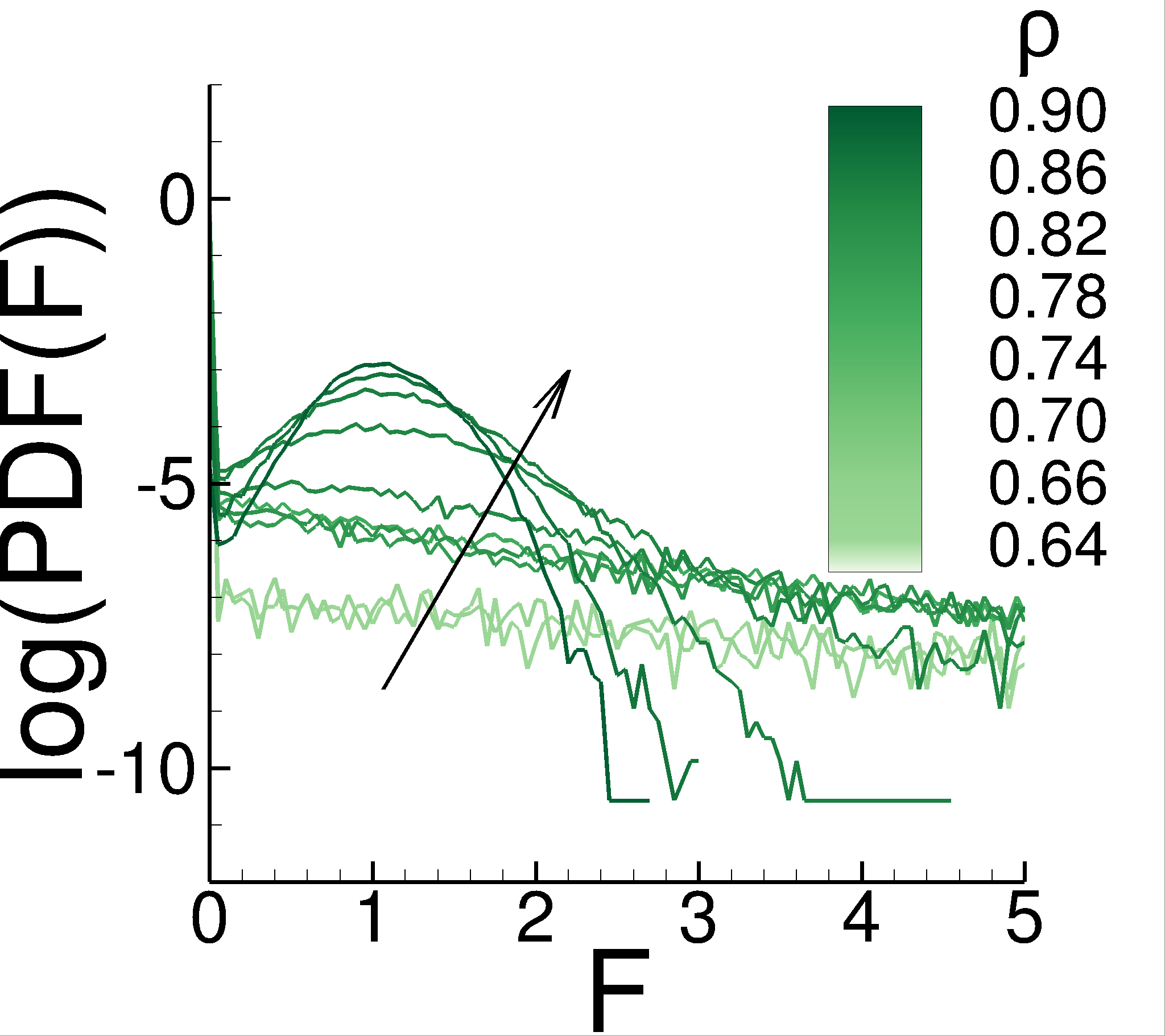}}\\
  \subfloat[]{\includegraphics[width=1.75in]{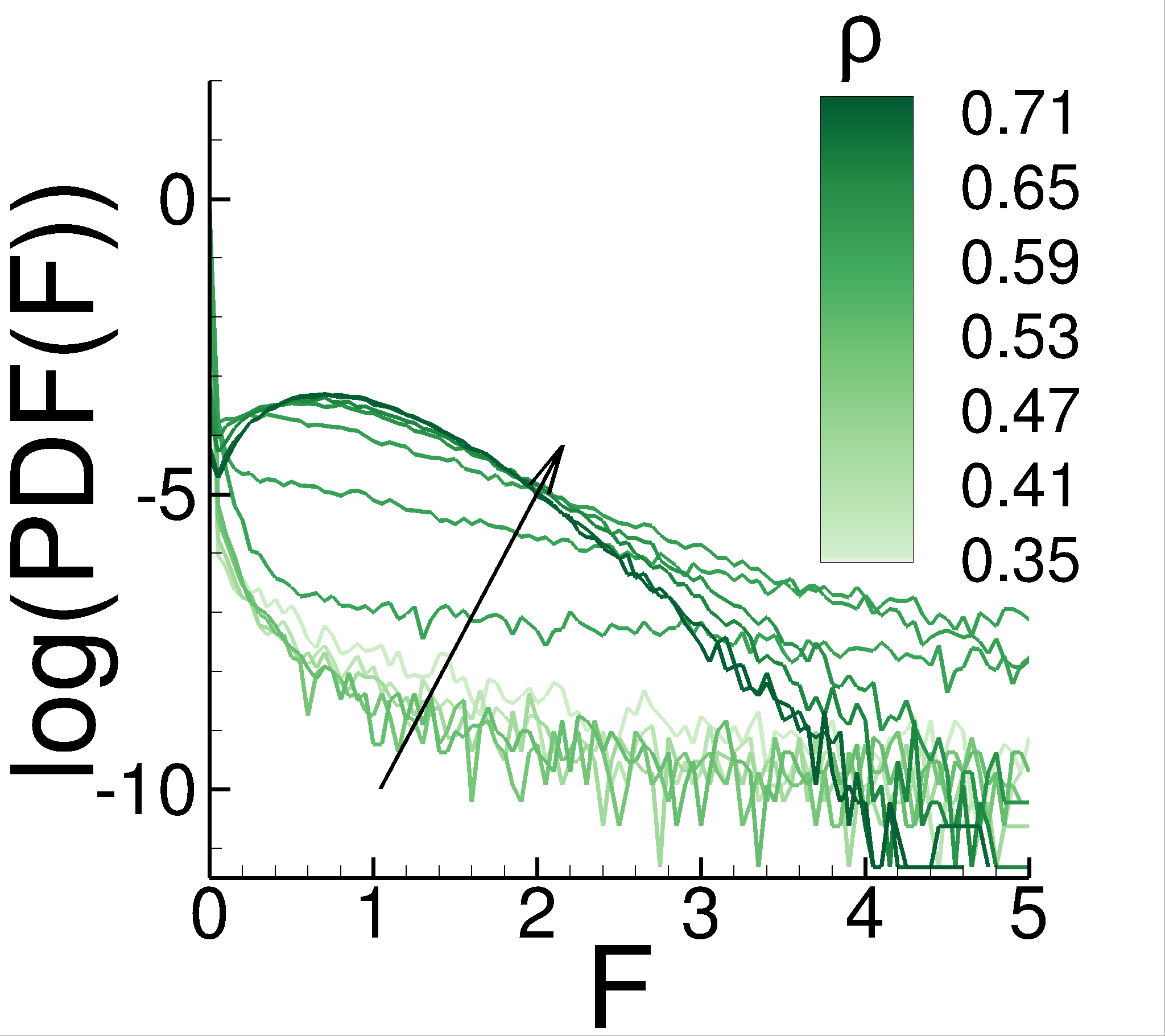}}
  \caption{PDF of forces for different values of $\rho$ 
  for (a) $2$D non-linear force model,  (b) 2D linear force model and (c) $3$D system.  
   The arrows indicate the direction of increased $\rho$. 
  }\label{fig:PDF}
\end{figure}

\subsection{Force Distribution}

Figure~\ref{fig:PDF} shows the force distribution, $\rm PDF(F)$, for the 
three considered systems.   
We note that the width of the distributions for the largest values 
$\rho\approx \rho_{\rm max}$ differs for the linear and two non-linear systems, with smaller width of $\rm PDF(F)$
in the linear system, suggesting a smaller variation of the forces and a more homogeneous
interaction network. The systems based on the non-linear force model have wider distribution and
in the $3$D system we observe the largest variation.   While the functional form of PDF(F) has been
widely discussed in the literature~\cite{brujic_2003, Minh2014, brodu_pre_15, mueth98, radjai_prl_96,azema_pre_07,Pugnaloni_2016,makse_prl_00}, we are not aware of the precise comparison of these functional forms between the systems characterized by different 
number of physical dimensions, or by different interaction force model.  

In the next section, we will use PDF(F) results together with the 
topological measures that will be introduced to discuss further distinguishing features of the interaction networks as the number of 
physical dimensions and the force model are varied.

\subsection{Force cluster analysis: non-percolating clusters}
\label{sec:non-percolating}

We continue our analysis of the forces between particles by examining the cluster size
distribution in the interaction networks. We omit the percolating cluster if it is present in the interaction network.
Since percolating cluster has a different characteristic scaling behavior~\cite{stauffer} than the average cluster size of the
non-percolating clusters~\cite{ostojic06, kovalcinova_scaling}, we 
discuss its properties separately. 

\begin{figure}[ht!]
  \centering
  \subfloat[]{\includegraphics[width=1.75in]{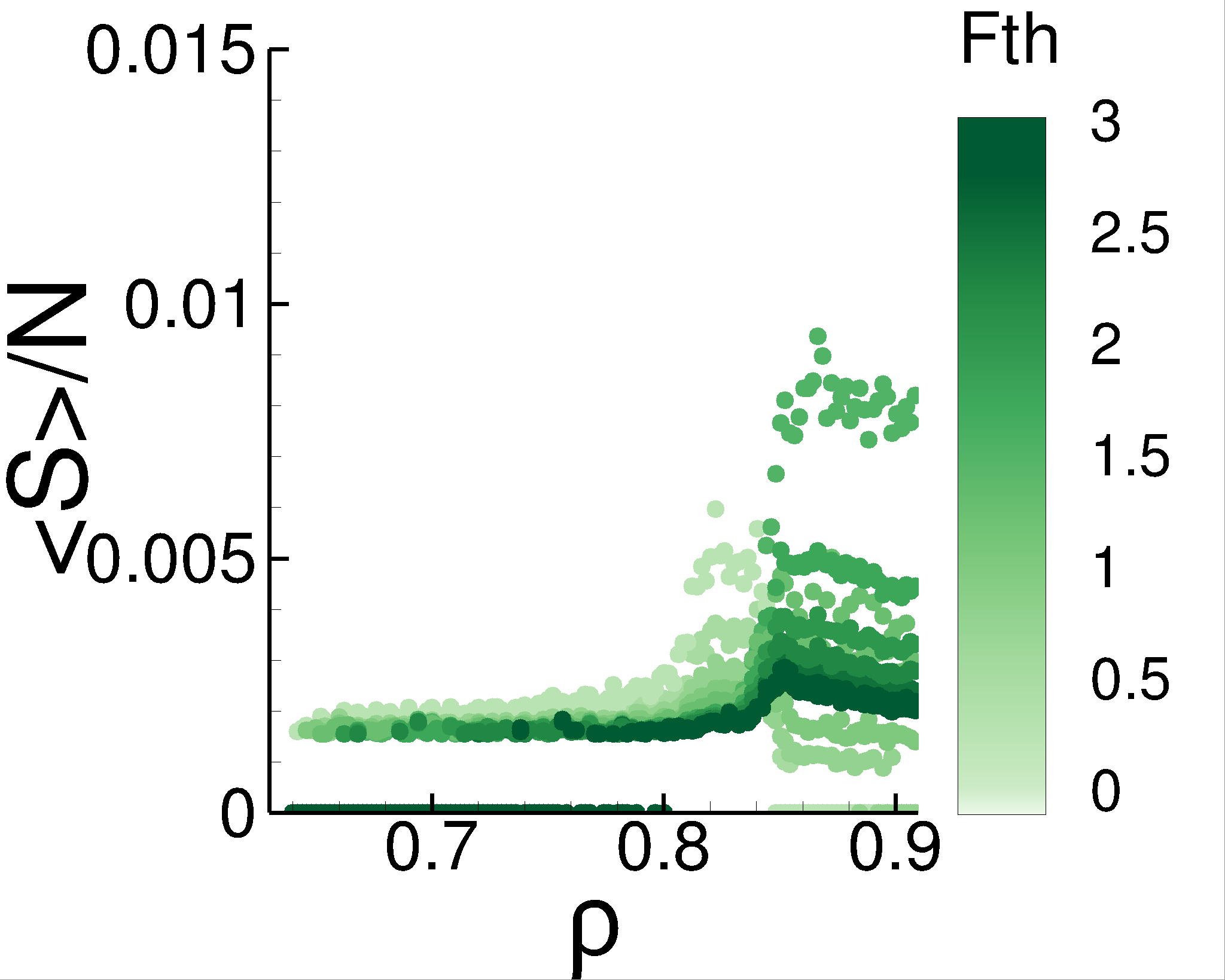}}
  \subfloat[]{\includegraphics[width=1.75in]{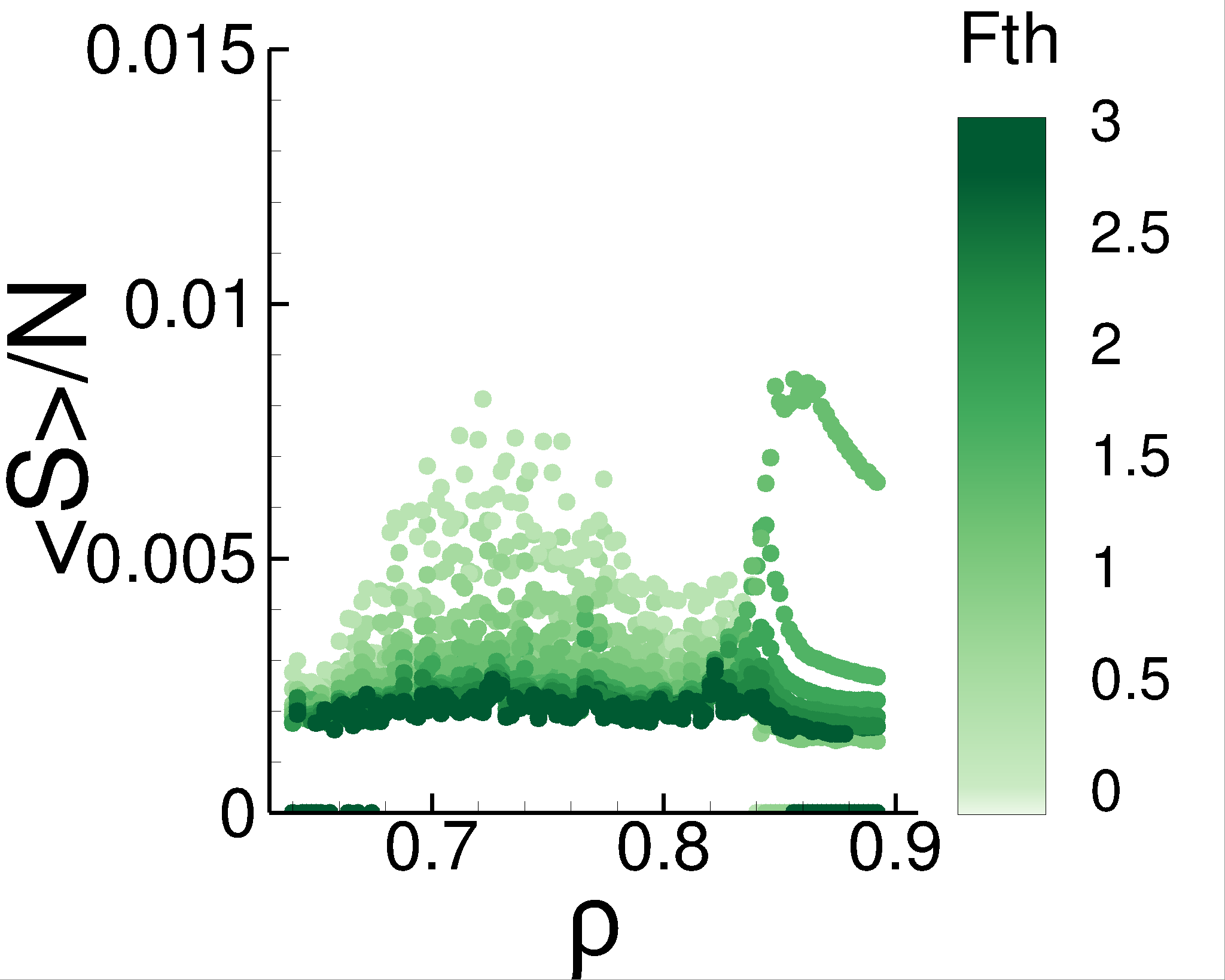}}\\
  \subfloat[]{\includegraphics[width=1.75in]{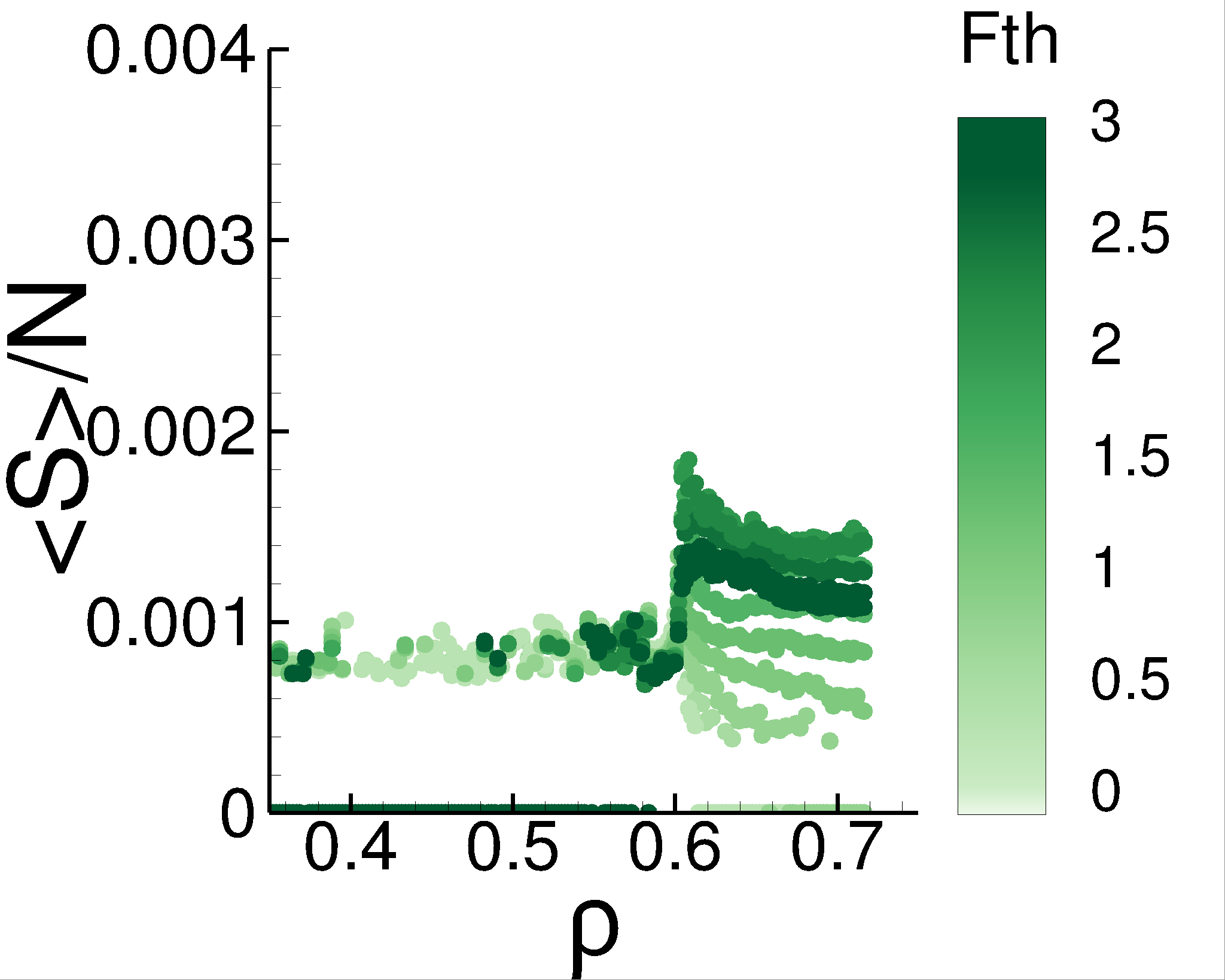}}
  \caption{Average cluster size, $\langle S\rangle$, normalized by the total number of particles, $N$, 
  for (a) $2$D non-linear force model,  (b) 2D linear force model and (c) $3$D system.  
}\label{fig:avg_S}
\end{figure}

 Figure~\ref{fig:avg_S} shows the average cluster size, $\langle S\rangle$, rescaled by the total number of particles, $N$, 
 for different values of $F_{\rm th}$.
In the $2$D systems, the largest values of $\langle S\rangle/N$ are observed for $F_{\rm th}=1.25$ and $F_{\rm th}=1.5$ for the linear and
non-linear systems, respectively. This finding implies that regardless of the force model, the interaction network is dominated by the force clusters composed of forces near the average force in $2$D.
Force model becomes important in $2$D when $F_{\rm th}\geq 2.0$ and $\rho>\rho_J$. Specifically, in the linear case, $\langle S\rangle/N$ decays faster than in the non-linear one.  In other words, the interaction network is more uniform for the linear system and we expect to observe a smaller variation of forces for $\rho\rightarrow
\rho_{\rm max}$. Indeed, our discussion of PDF's in the context of Fig.~\ref{fig:PDF} confirms that the $\rm PDF (F)$ in the linear system follows
a more narrow distribution than the $\rm PDF (F)$ in the non-linear one.

Turning now our attention to the influence of the number of physical dimensions, we 
compare the panels (a) and (c) of Fig.~\ref{fig:avg_S}.  The most obvious
difference between the two  is the
$F_{\rm th}$ for which $\langle S\rangle/N$ attains the maximum for $\rho>\rho_J$.
In the $3$D system, the largest values of $\langle S\rangle/N$ are observed for 
$F_{\rm th}\approx 2.0$. Therefore, the interaction networks in $3$D contain more 
clusters with large forces - the forces are less concentrated
around the mean force. The $\rm PDF(F)$ for the 3D system shown in Fig.~\ref{fig:PDF} is indeed wider
confirming larger force variation in comparison to $2$D.

\subsection{Force cluster analysis: percolating cluster}

\begin{figure}[ht!]
  \centering
  \subfloat[]{\includegraphics[width=1.75in]{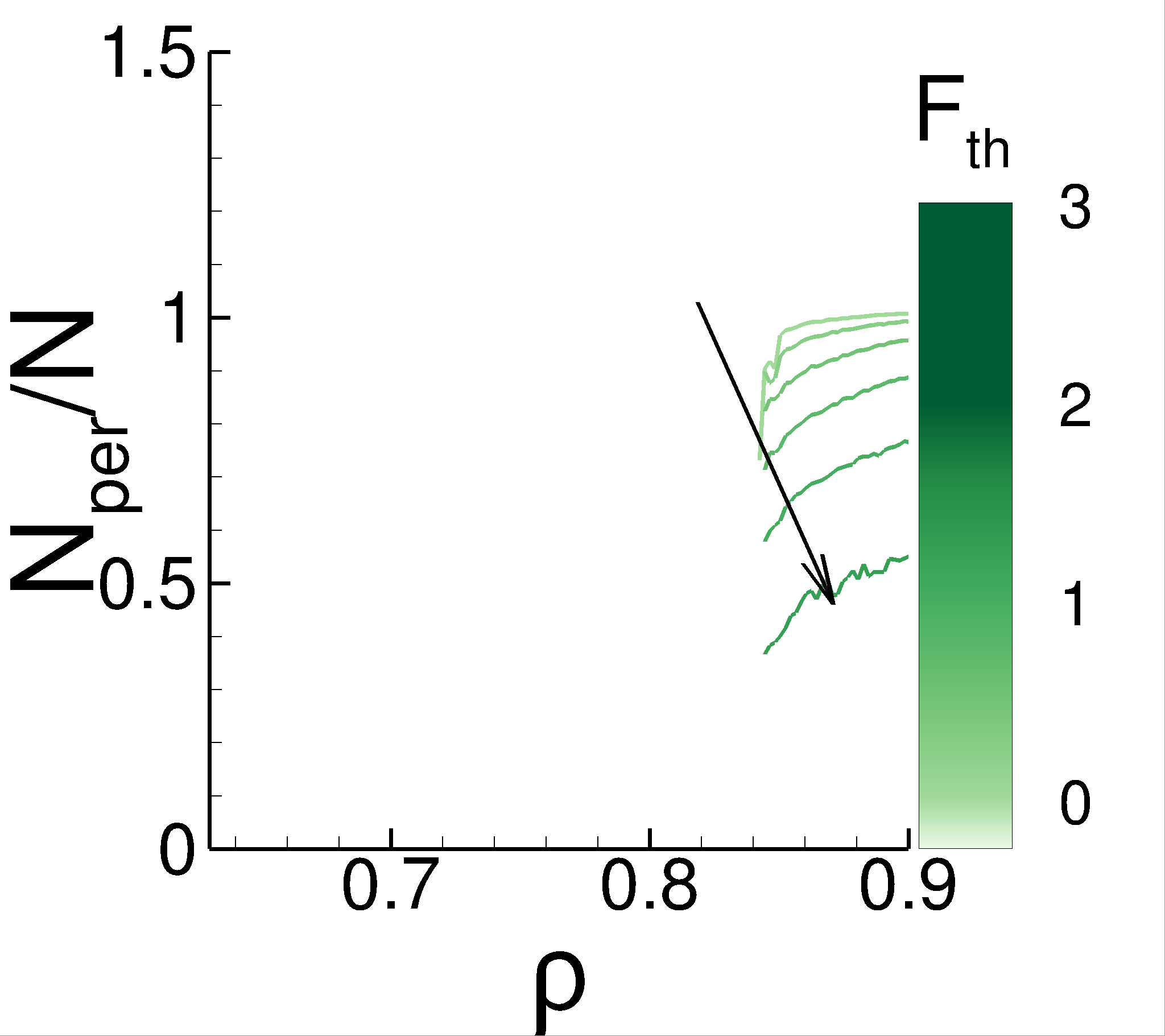}}
  \subfloat[]{\includegraphics[width=1.75in]{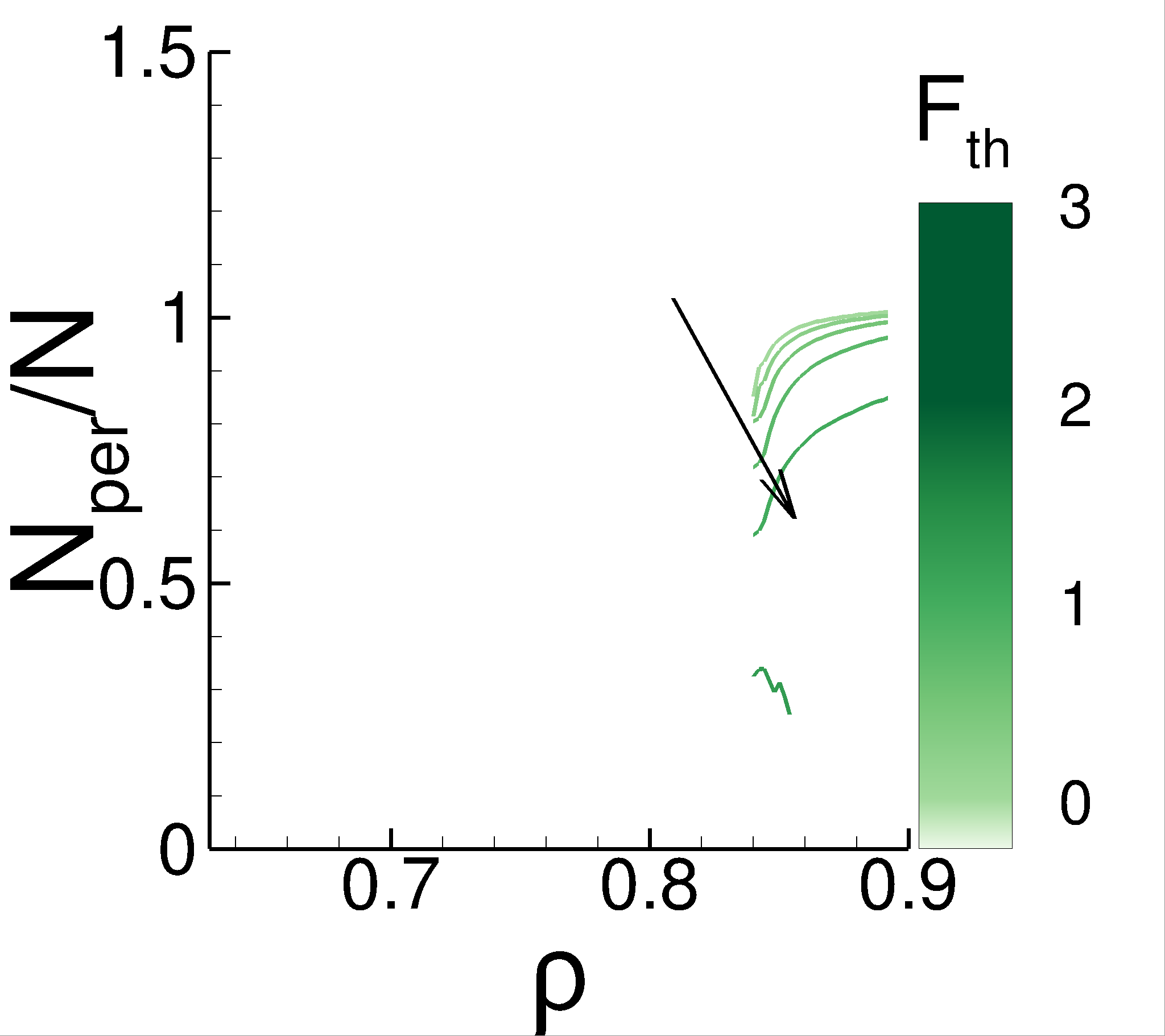}}\\
  \subfloat[]{\includegraphics[width=1.75in]{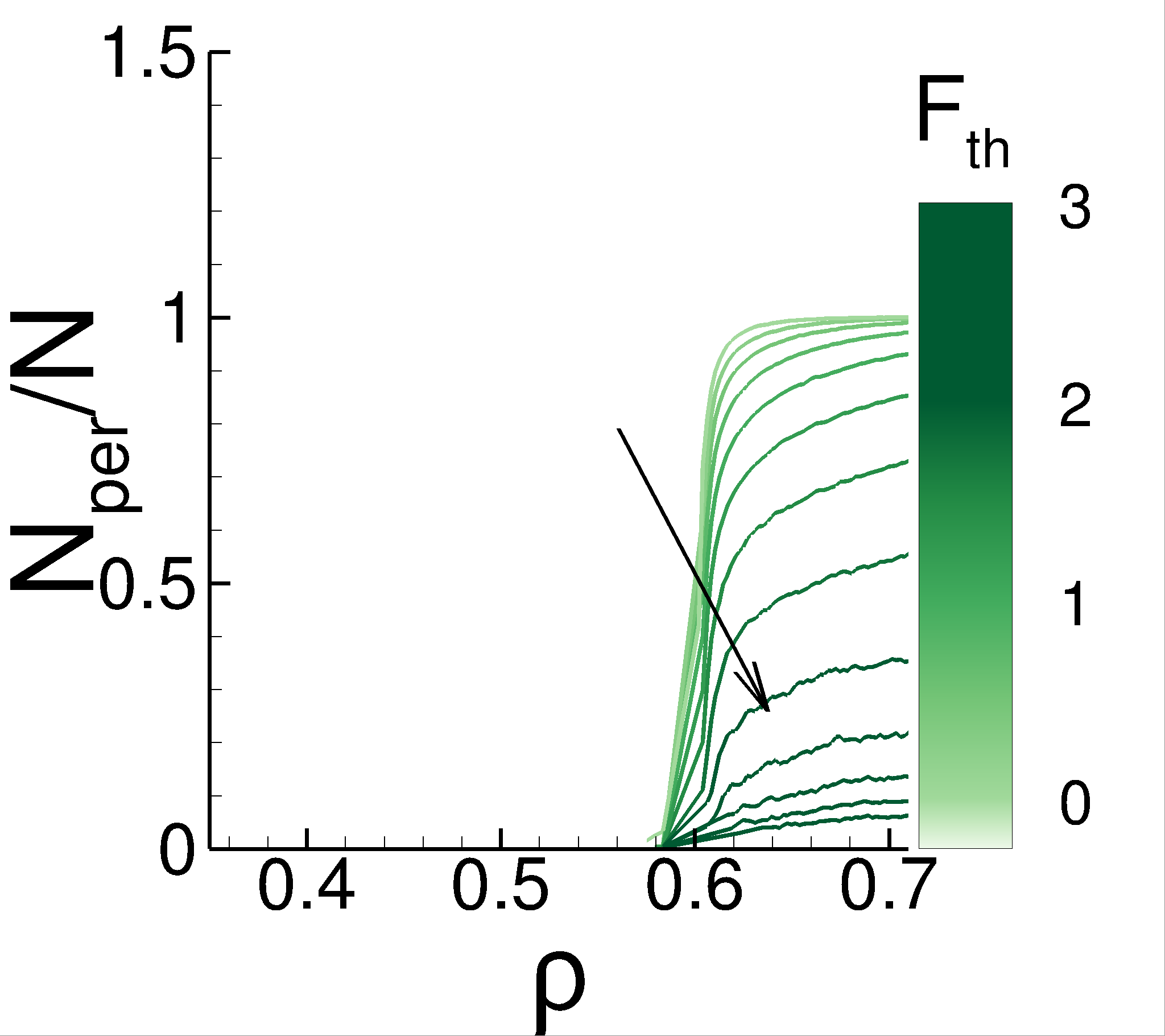}}
  \caption{Proportion of the particles in a percolating cluster 
  for (a) $2$D non-linear force model,  (b) 2D linear force model and (c) $3$D system  
   for different values of $F_{\rm th}$. 
    The arrows indicate the direction of increased $F_{\rm th}$.   
   }\label{percolation_info}
\end{figure}
Figure~\ref{percolation_info} shows the average force and the amount of particles participating in a percolating
cluster, $N_{\rm per}$, normalized by the total number of particles, $N$.  We remind the reader that 
the results are averaged over $20$ realizations for each system; here we only show the results if 
at least $50\%$ of the realizations contain a percolating cluster.
Recall that percolation is observed only when the system jams, consistently
with our previous findings~\cite{kovalcinova_percolation}.
We note that when we take into account all the contacts, i.e. when $F_{\rm th} \approx 0.0$,
 the percolating cluster in the $2$D systems contains almost
all the particles. However, in the $3$D case, in particular for $\rho\approx\rho_J$,  the percolating cluster contains only a very small number of particles, as it can be seen 
in Fig.~\ref{percolation_info}.  

The reason why  less particles are needed for percolation in $3$D is purely geometrical and is discussed next.
We consider the interaction networks at $F_{\rm th } = 0.0$ first and represent them as random networks with
particles as vertices and contacts represented by edges.
We also define the mean degree of a random network, $\langle k\rangle$
as the mean number of edges per vertex (in the context of granular systems $\langle k\rangle$
is equivalent to the average contact number, $Z$).
Let us now assume that a given random network has $N$ vertices and
a mean degree $\langle k\rangle$ and ask what is the maximum number of vertices
we can randomly remove and still have a percolating cluster.
From the theory of random networks~\cite{cohen_2004, Newman_2010}, a critical occupation probability
is estimated by
\begin{equation}
  \phi_{\rm c} = {\langle k\rangle \over \langle k^2\rangle - \langle k\rangle}\label{eq:occ_ratio}\,.
\end{equation}
($\phi_{\rm c}$ is the smallest
ratio of the vertices present in a random network leading to a percolating cluster).
Note that $\phi_{\rm c}$ provides a lower bound for $N_{\rm per}/N$.
We can now use Eq.~(\ref{eq:occ_ratio})
to estimate $\phi_{\rm c} $ in $2$D and $3$D.
In our simulations close to jamming, we find that, as expected, $\langle k\rangle \approx 3$ and $\langle k\rangle \approx 6$
in $2$D and $3$D, respectively.   Note that the exact value of $\langle k\rangle$ depends on the friction between 
particles~\cite{Liu, hecke_2010, Henkes_epl_2010, majmudar07a} and is estimated by $\langle k \rangle \approx d+1 + (2N_{\rm m}/d)$
where $N_{\rm m}$ is the mean number of contacts that have tangential forces equal to the Coulomb 
threshold~\cite{Papadopoulos_18, Shundyak_pre_2007, Henkes_epl_2010, Liu}. If we assume that $\langle k^2\rangle\approx\langle k\rangle^2$ (this assumption would be exactly satisfied if each particle had the same number of contacts), 
we find $\phi_{\rm c} \simeq 0.5$ in $2$D and 
$\phi_{\rm c} \simeq 0.2$ in $3$D at $\rho_J$. This estimate provides intuitive (even if only approximate) explanation of why we observe smaller (relative to the 
total number of particles) percolating clusters in $3$D compared to $2$D systems when $\rho \approx \rho_J$. Note that
this argument is purely geometrical and does not depend on simulation protocol, particle properties (other than friction
that plays a role in the $\langle k\rangle$ estimate), or the system size.

 Our results show that the differences in $N_{\rm per}/N$ extend to the non-zero force thresholds and
 $\rho > \rho_{\rm J}$. 
 Specifically, we find that whenever percolating cluster exists for the $2$D systems,
it is composed of a significant number of particles. In $3$D, on the
other hand, we can find a percolating cluster composed of a very small percentage of particles, particularly
when $F_{\rm th}>2.0$ or when $\rho\approx\rho_{\rm J}$. Let us also reiterate that the presented results 
are robust with the respect to the details of the relaxation period, at least within the range that we 
could consider using available computational resources.

\section{Topology of interaction networks}
\label{sec:topology}

We now continue with the study of topological properties of interaction networks.
We focus on the number of components/clusters in the interaction network, the number of loops and 
voids (to be defined below) and on the measures emerging from persistent homology~\cite{pre13}.   
The general motivation is that Betti numbers and persistence analysis can be used to compare 
the geometries of the force distributions~\cite{kramar_2014} and find their characteristic behavior
during the jamming transition~\cite{kramar_pre_14}.
We will see in what follows that this type of analysis indeed allows for quantifying 
significant differences between the topology of interaction networks in $2$D and $3$D systems, as
well as in $2$D systems characterized by different interaction force models.
To start with, we discuss the simplest topological measure, Betti numbers.

\subsection{Betti numbers}
\label{sec:betti}

Zeroth Betti numbers, $B_0$, denotes the number of
components (clusters), and $B_1$ denotes the number of loops in a force
network (a loop in an interaction network is a collection of connected edges forming a
closed cycle). The next Betti number, $B_2$, relevant to $3$D geometry, counts the number of cavities.   Note that $B_0$ provides
different (but related) information than the average cluster size, $<S>$, discussed in
Sec.~\ref{sec:non-percolating}; $B_0$ provides the information about 
the number of clusters, independently of their size.

When analyzing Betti numbers, $B_1$ in particular, one has to make a choice of whether to consider the loops made out of three particles
in contact, sometimes referred to as 3-cycles in the studies of interaction networks in granular systems~\cite{tordesillas_pre10}.  
 It is known that such loops play a role in 
 stability of interaction networks in granular systems~\cite{tordesillas_pre10,ardanza_pre14}. 
 However, since 3-cycles are 
 prevalent and not unique to a specific set of physical parameters or number of physical dimensions, we ignore them
in the present analysis and focus on the loops involving at least four particles, similarly as it
was done in our earlier studies~\cite{kramar_2014,pre13, kramar_pre_14, epl12}.

\begin{figure}[ht!]
  \centering
  \subfloat[]{\includegraphics[width=1.7in]{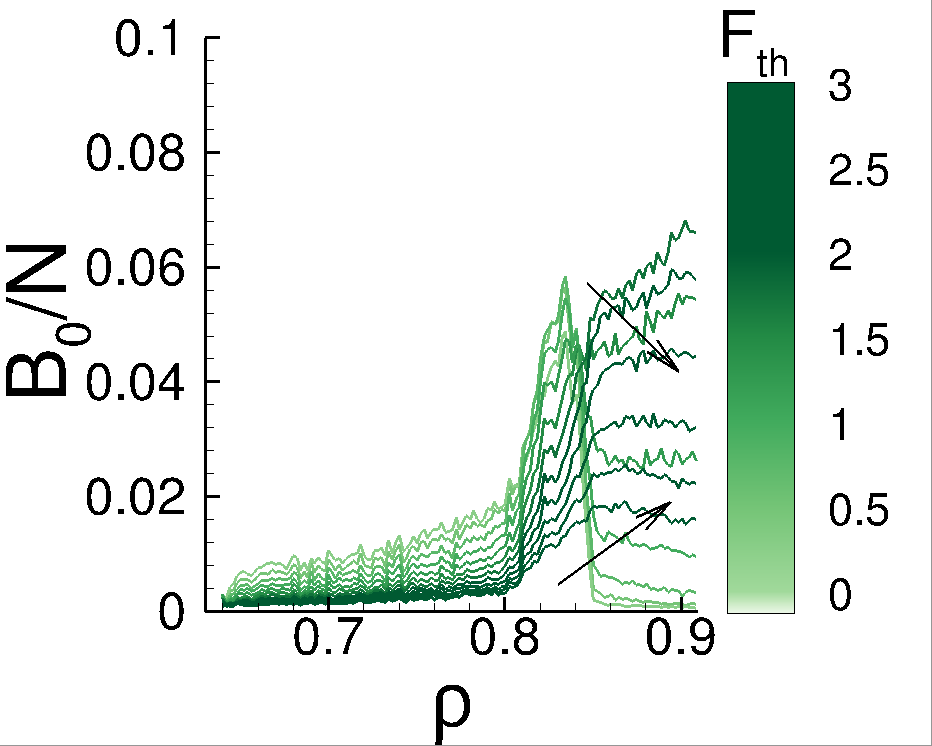}}
  \subfloat[]{\includegraphics[width=1.7in]{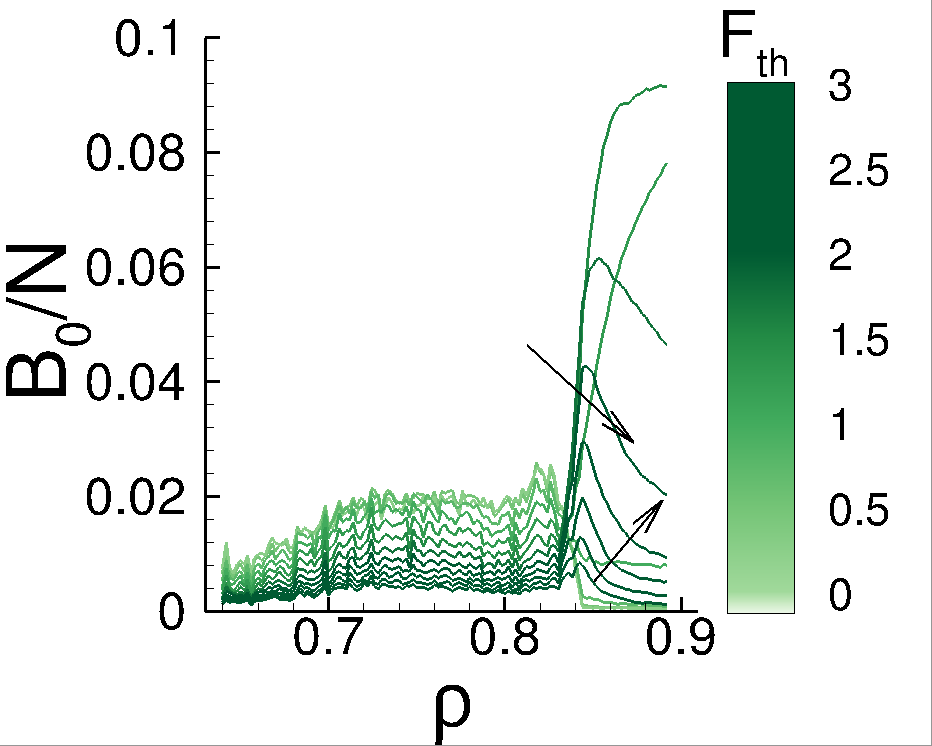}}\\
  \subfloat[]{\includegraphics[width=1.7in]{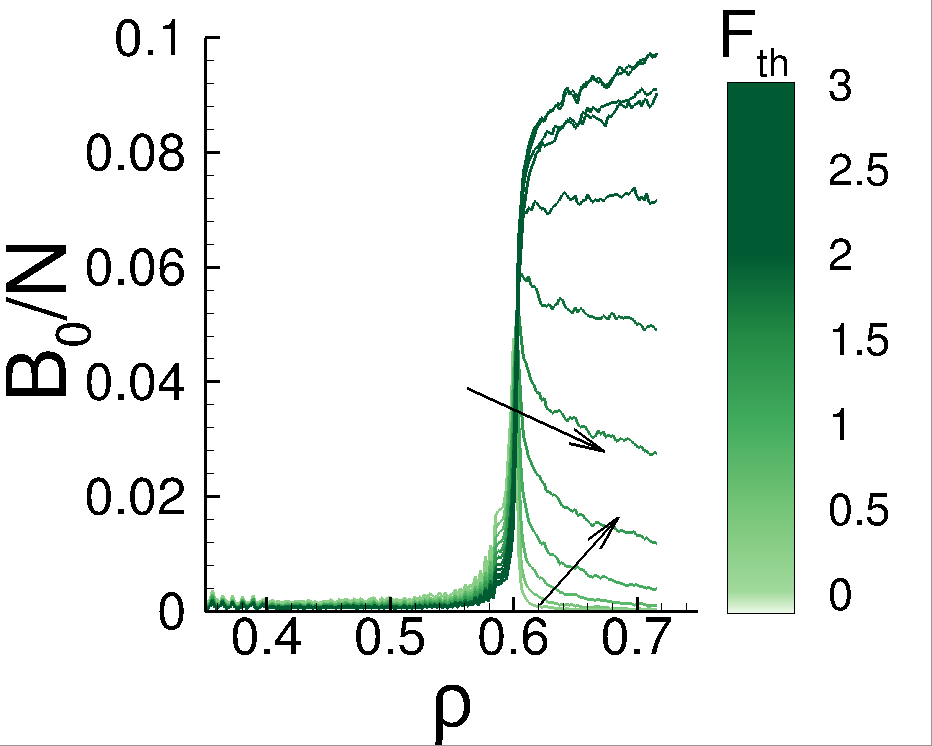}}
  \subfloat[]{\includegraphics[width=1.7in]{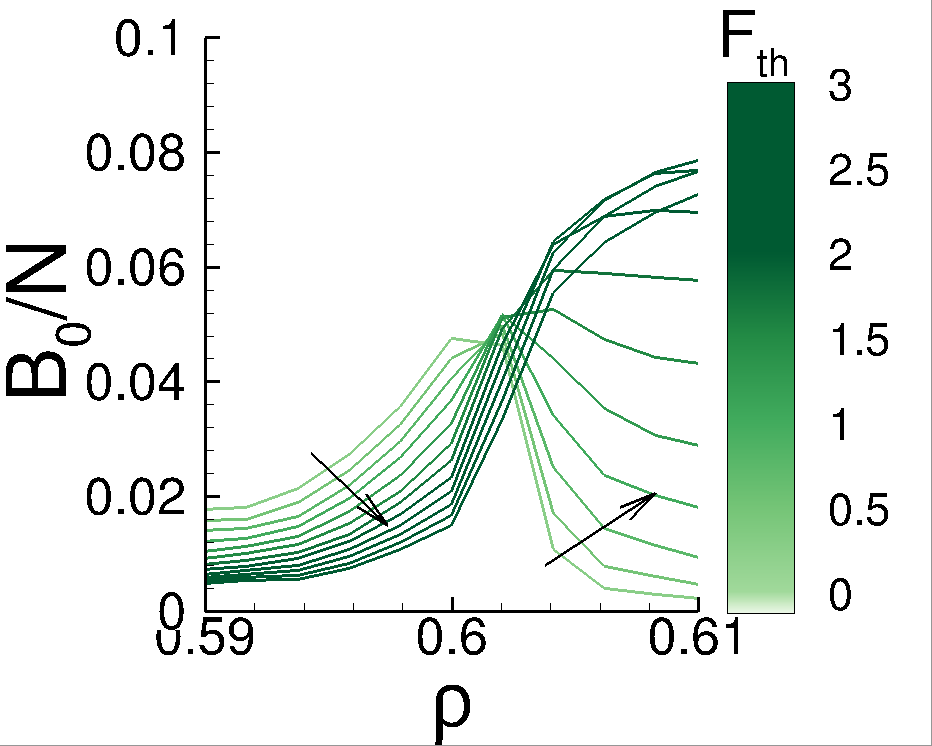}}
  \caption{Average number of clusters/components, $B_0$, rescaled by the total number of  particles, $N$,
   for (a) $2$D non-linear force model,  (b) 2D linear force model and (c) $3$D system.  The panel (d) is a 
   closeup of the $3$D granular system showing a clear peak in $B_0/N$ for small $F_{\rm th}$. 
      In (a~-~c) the arrows pointing up show an (increasing) trend of the curves from the smallest
 $F_{\rm th}$ and then the arrows pointing down show 
  a (decreasing) trend  when $F_{\rm th}\rightarrow \max\{F_{\rm th}\}$.}\label{fig:B_0}
\end{figure}

Let us begin by examining the number of clusters, $B_0$, forming in the interaction network
during compression. For a complete representation of the interaction network evolution, a
range of force thresholds is considered for each packing fraction.
Figure~\ref{fig:B_0}, which shows a set of cross-sections of the introductory 
Fig.~\ref{fig:B0}, plots $B_0$ rescaled by the total number of particles, $N$.
When considering Betti numbers, inclusion of percolation cluster is not relevant, 
since it modifies $B_0$ only by unity; for simplicity we include it in the Betti number
count. 

Figure~\ref{fig:B_0}(a),~(c) shows that for small values of $F_{\rm th}<0.5$,
the systems based on the non-linear force model develop a pronounced peak near the
respective jamming packing fractions, $\rho_{\rm J}$ (see in particular Fig.~\ref{fig:B_0}(d), which is a zoom-in of 
Fig.~\ref{fig:B_0}(c)).

\begin{figure}[ht!]
  \subfloat[]{\includegraphics[width=1.7in]{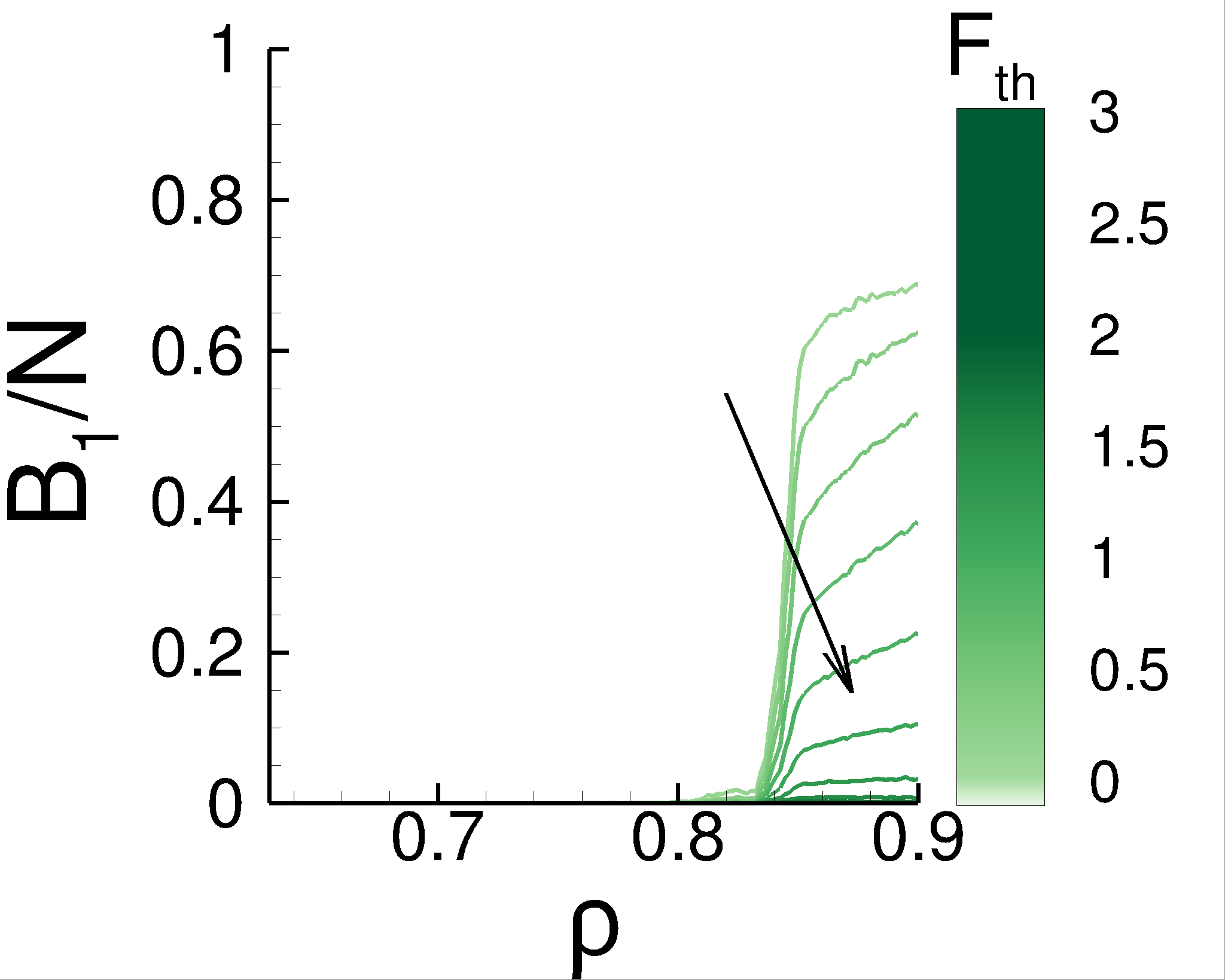}}
  \subfloat[]{\includegraphics[width=1.7in]{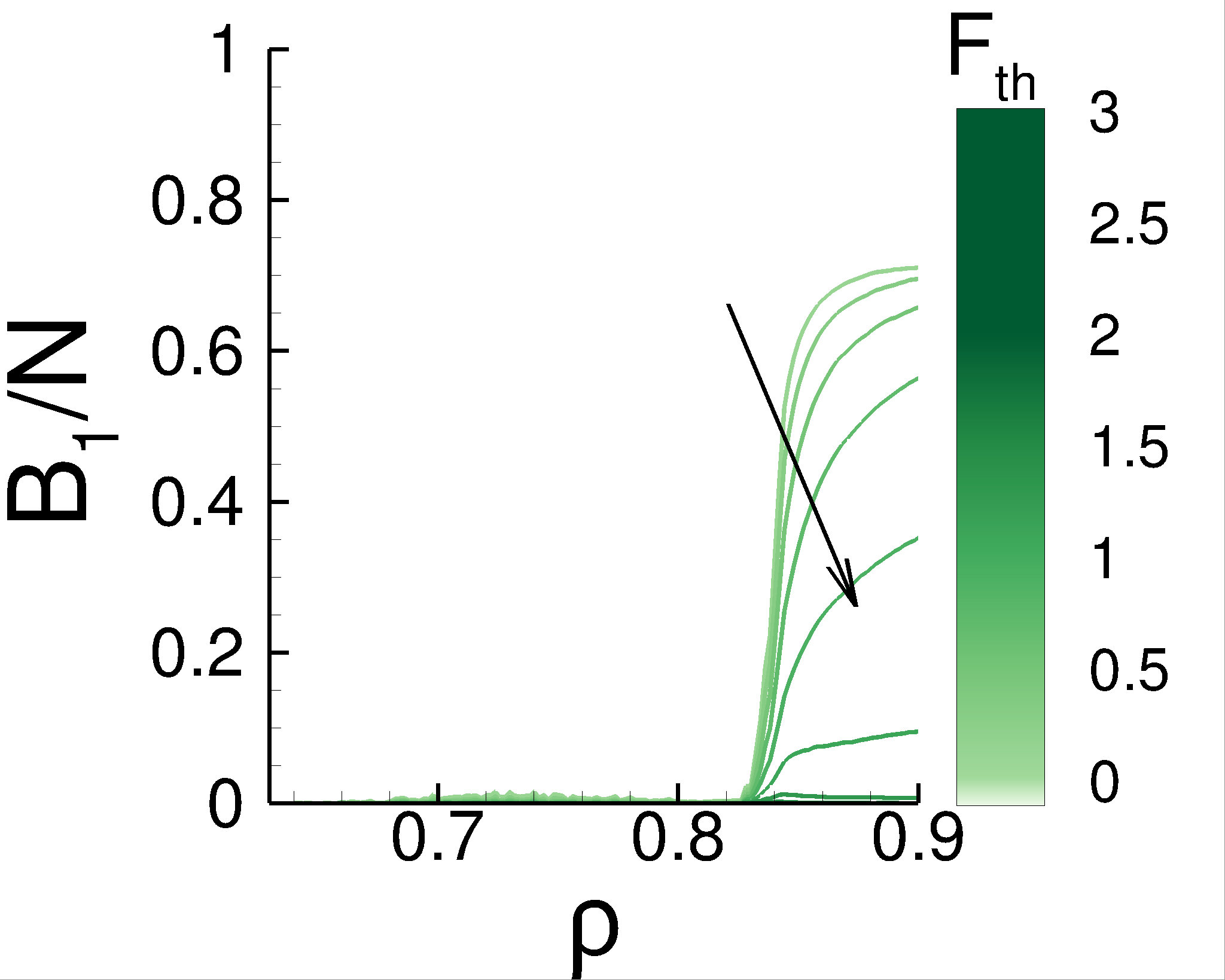}}\\
  \subfloat[]{\includegraphics[width=1.7in]{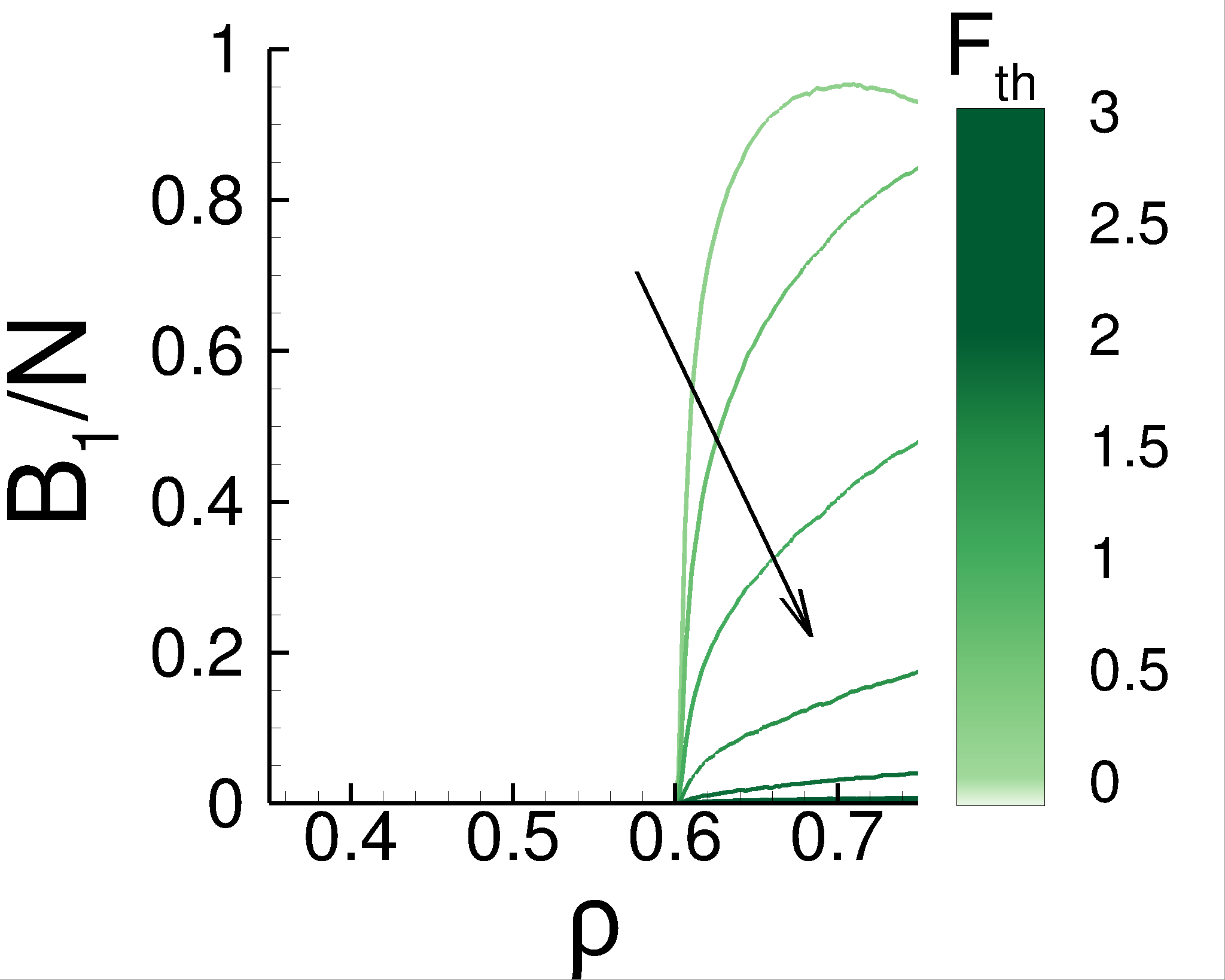}}
  \caption{The evolution of the number of loops, $B_1$, normalized  by the total number of
  particles, $N$, during compression
    for (a) $2$D non-linear force model,  (b) 2D linear force model and (c) $3$D system.   
   The arrows indicate the direction of increased $F_{\rm th}$.  
  }\label{fig:B_1}
\end{figure}

The observations made based on  Fig.~\ref{fig:avg_S}
suggest that there are important structural differences
between the linear and non-linear systems.  For the 
non-linear force model, we find that for $F_{\rm th}<0.5$ and $\rho\approx \rho_J$
the clusters are more numerous 
compared to the 
ones found for  $\rho > \rho_J$ (leading to the peaks visible in 
Figs.~\ref{fig:B_0}(a), (c), (d)).  
In contrast, for the linear force model 
the peak in $B_0/N$ is rather insignificant.

\begin{figure}[ht!]
  \subfloat[]{\includegraphics[width=1.7in]{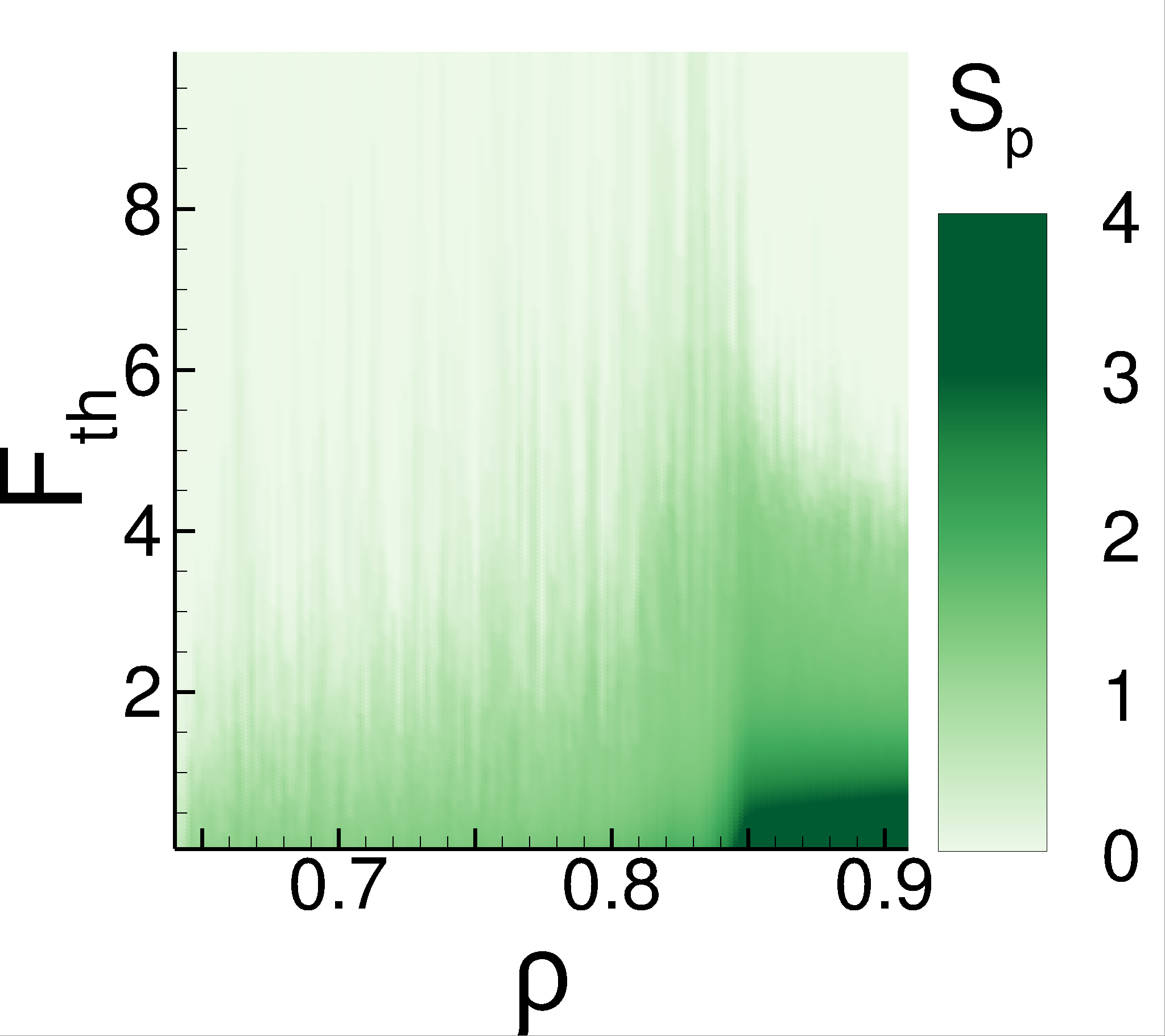}}
  \subfloat[]{\includegraphics[width=1.7in]{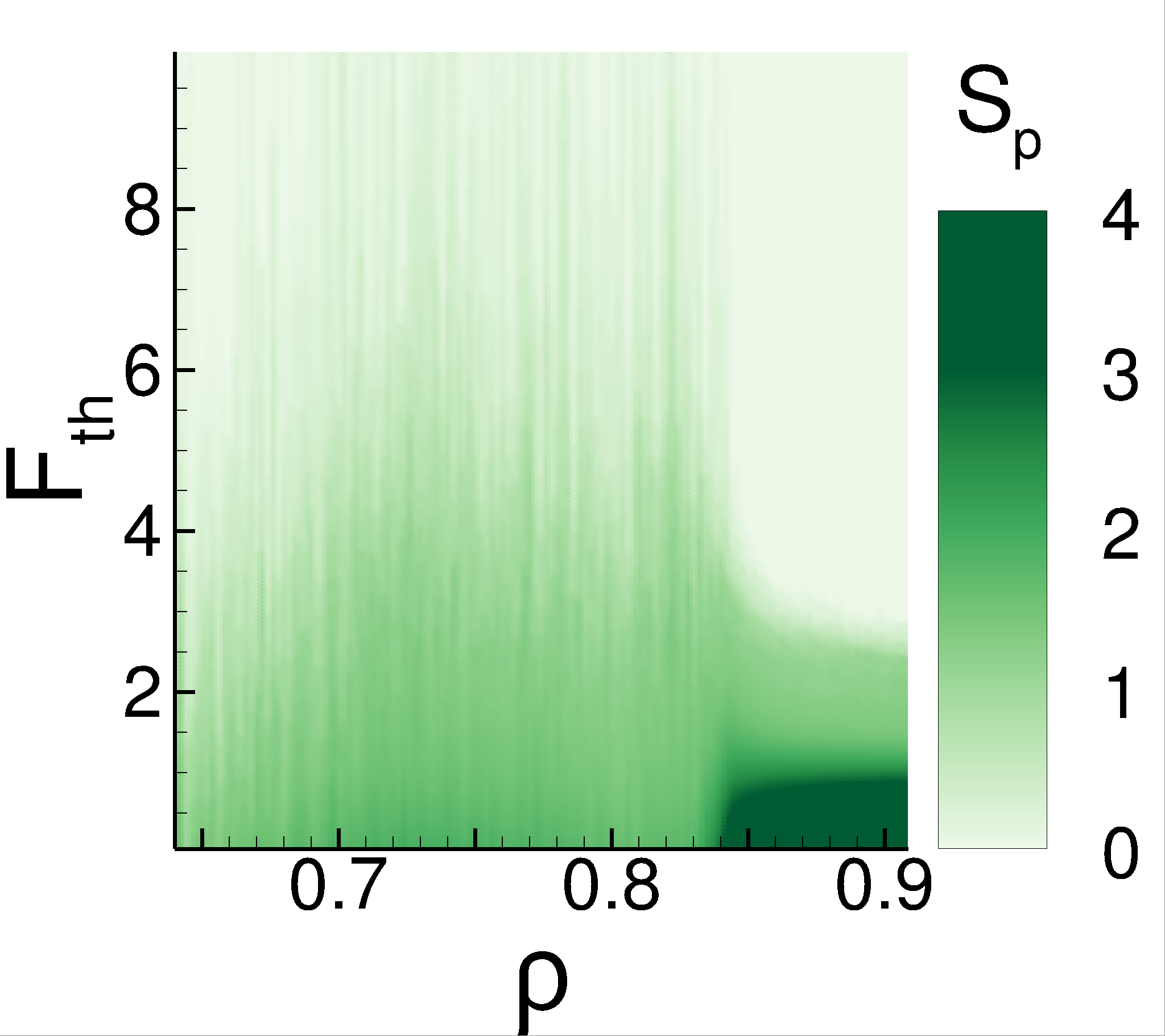}}\\
  \subfloat[]{\includegraphics[width=1.7in]{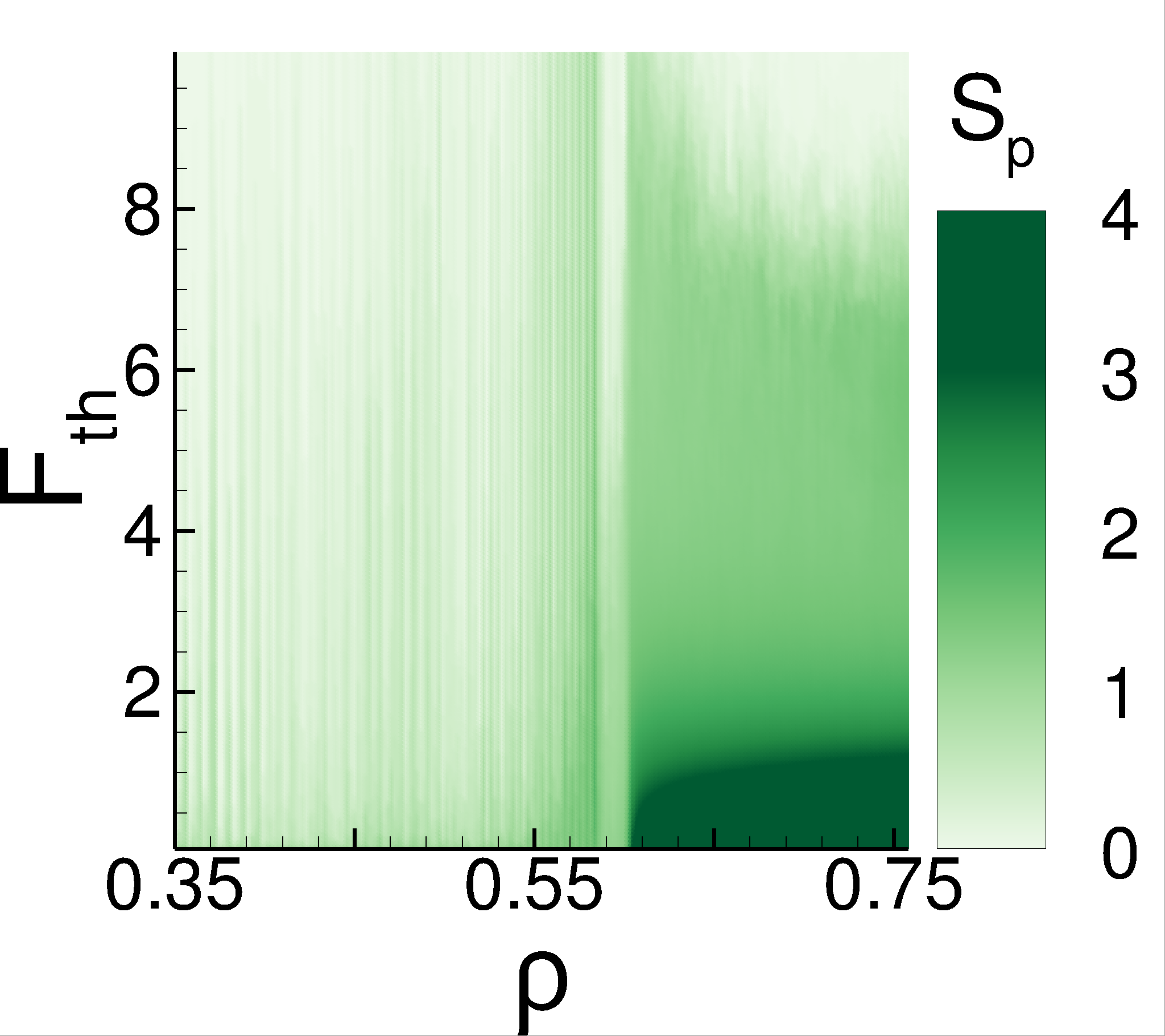}}
  \caption{Shape parameter, $S_{\rm p}$, as a function of $\rho$ for all values of $F_{\rm th}$ 
    for (a) $2$D non-linear force model,  (b) 2D linear force model and (c) $3$D system.   
  }\label{fig:shape}
\end{figure}

\begin{figure}[ht!]
  \subfloat[]{\includegraphics[width=1.7in]{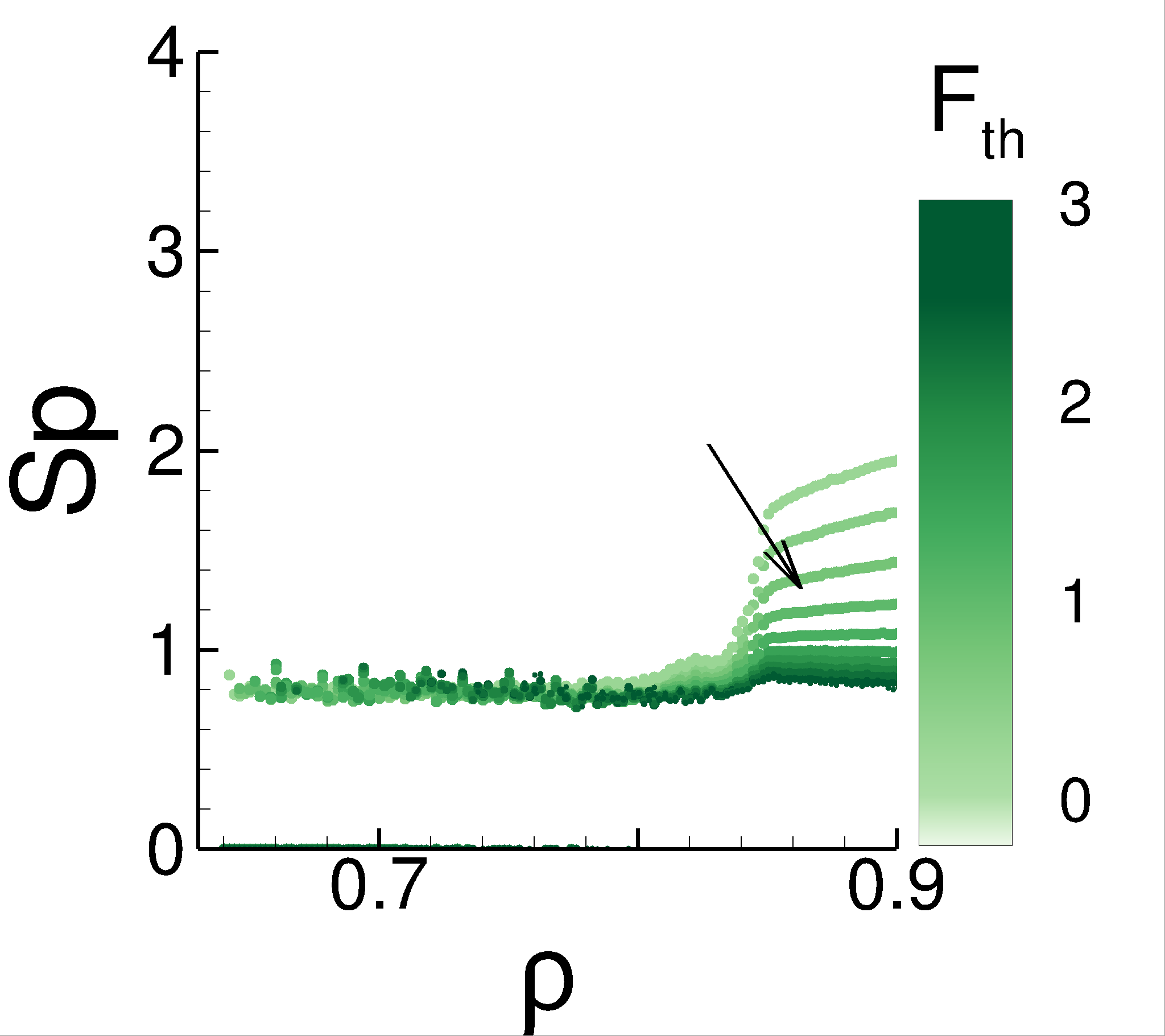}}
  \subfloat[]{\includegraphics[width=1.7in]{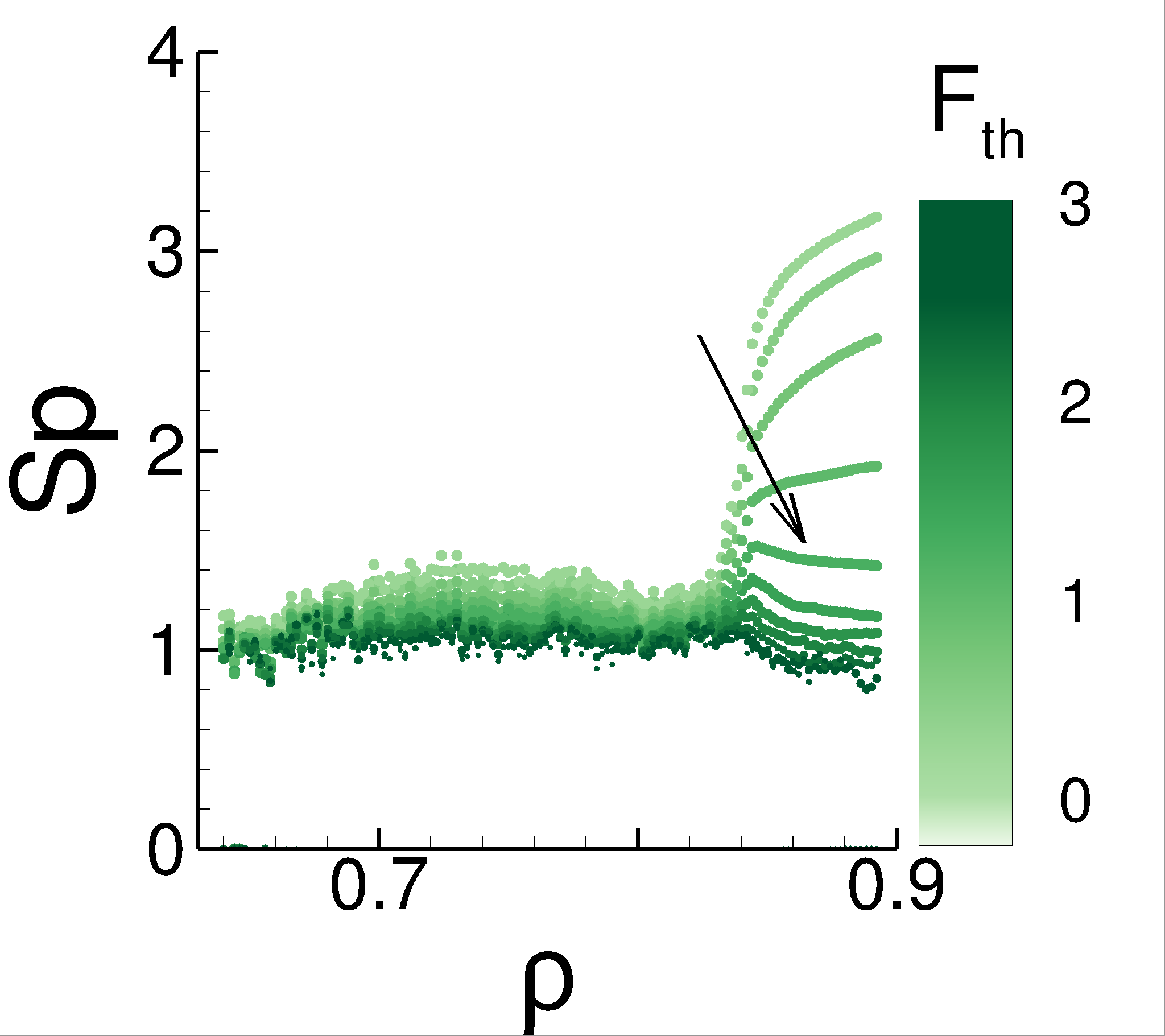}}\\
  \subfloat[]{\includegraphics[width=1.7in]{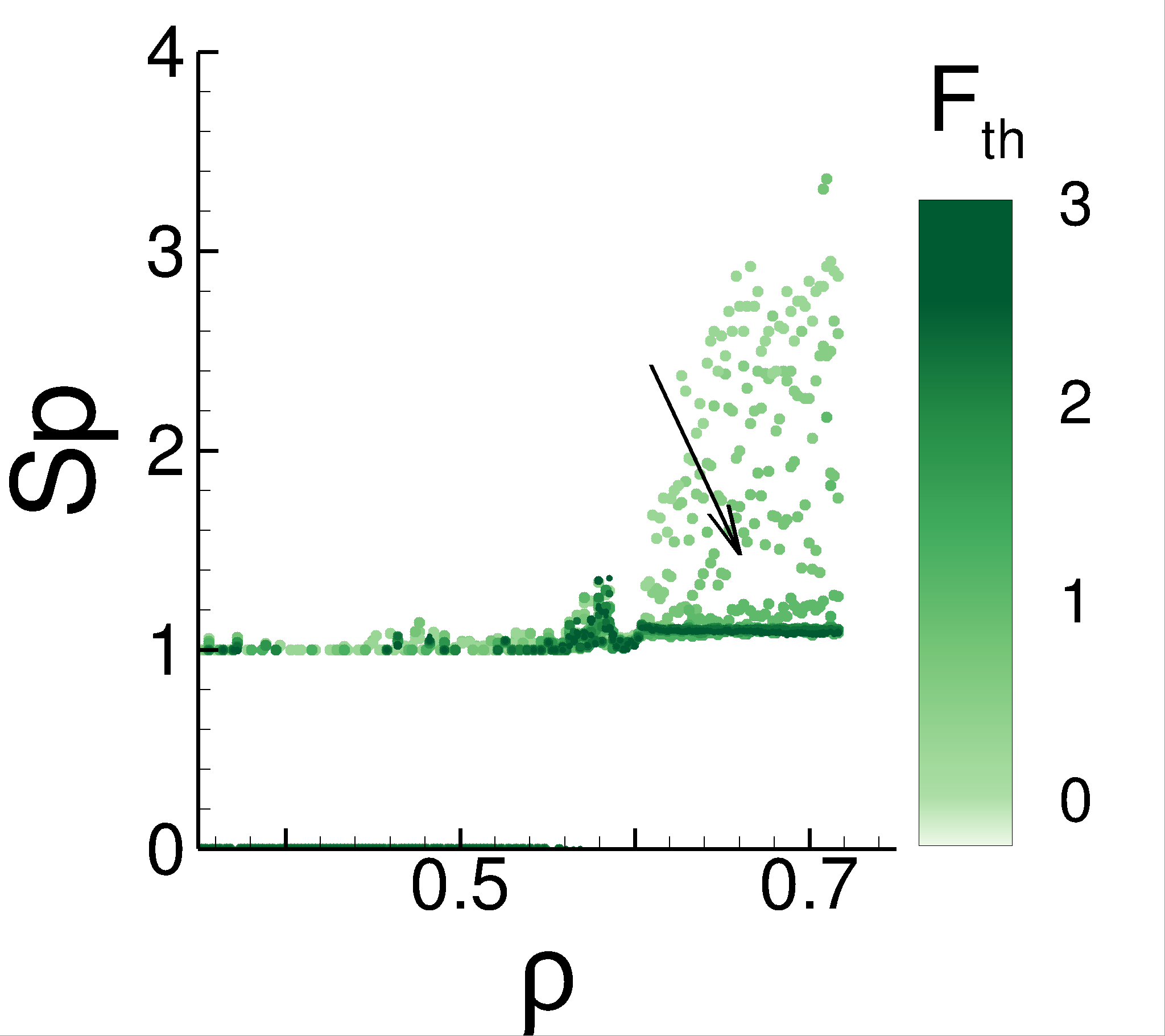}}
  \caption{Shape parameter, $S_{\rm p}$, as a function of $\rho$ for the chosen values of $F_{\rm th}$
   for (a) $2$D non-linear force model,  (b) 2D linear force model and (c) $3$D system.    
  The arrows indicate the direction of increased $F_{\rm th}$.
}\label{fig:shape_selected}
\end{figure}

Figure~\ref{fig:B_1} shows, perhaps surprisingly, that $B_1/N$ results follow
the same trend for all systems considered.   The $B_1/N$ evolution is generic during
compression: for small $\rho$ below jamming
we do not observe any loops forming and only beyond jamming, $B_1/N$ starts to rise and then plateaus
for $\rho$ close to $\rho_{\rm max}$. The only visible difference between $2$D and $3$D systems is the larger magnitude
of $B_1/N$ for the  $3$D case.  

The evolution of the next Betti number, $B_2/N$, is not analyzed here in detail  since
 $B_2$ is not defined in $2$D. We only briefly mention that in $3$D,
 $B_2$ vanishes for $\rho < \rho_{\rm J}$ and grows in a monotonous fashion as $\rho$ increases beyond $\rho_{\rm J}$.  

\subsection{Shape of the clusters}

The Betti numbers provide only the count of the clusters and loops in the interaction network
and do not provide any information about the cluster shape; based on Betti numbers, we do not
know whether the clusters are isotropic, chain-like, or of some other form.  
To explore the cluster shape, we introduce a shape parameter, $S_{\rm p}$, defined
as the number of edges in the clusters divided by the number of participating particles, 
averaged over all realizations.  The motivation behind such measure is the
following.  
If the particles that participate in a cluster all have one or
two contacts, the cluster forms either an open or closed path, defined as a sequence of vertices 
(here particles). Note, that a closed path is a loop and contributes to $B_1$ count.
Let us assume that the number of particles in such a path is $N_{\rm L}$. Then, the number of inter-particle
contacts forming an open path is $N_{\rm L} - 1$ and the resulting shape parameter for
such a path is $\bar S_{\rm p} = (N_{\rm L} - 1)/N_{\rm L}\in[0.5,1.0)$. For any loop, we have $\bar S_{\rm p}=1.0$.  
For more complex shape structures (not simple paths), 
we expect that $\bar S_{\rm p}\in (1.0,4.0]$ in $2$D and $\bar S_{\rm p}\in (1.0,7.0]$ in $3$D. (To avoid any confusion, we refer by $S_{\rm p}$ to the measure averaged over all clusters
and realizations, and by $\bar S_{\rm p}$ to the measure describing shape of an individual cluster, or to the measure
describing the clusters of an individual realization. )
The upper values of $\bar S_{\rm p}$  are estimated from the maximum
average contact number for the $2$D and $3$D systems.  For the purpose of simplifying
discussion below, we refer to the clusters with more than (on average) two contacts per
particle as complex clusters; the clusters that are not complex are simple.

Figure~\ref{fig:shape} shows $S_{\rm p}$ for all systems and for all values of
$F_{\rm th}$ and for varying packing fraction, $\rho$;
Figure~\ref{fig:shape_selected} shows cross-sections of $S_{\rm p}$ for selected values of $F_{\rm th}$.
Note that we put $S_{\rm p} = 0$ whenever $\bar S_{\rm p}=0$ for at least half of the
realizations. Otherwise the values are averaged only over the realizations such that $\bar S_{\rm p}\neq 0$.

In discussing the results shown on Figs.~\ref{fig:shape} and~\ref{fig:shape_selected}, we focus first on 
the 2D systems, and small packing fractions, $\rho< \rho_J$.   We observe that for both systems $S_{\rm p} \approx 1$.
Further insight is obtained by relating this result to 
 the ones for $B_1/N$ shown in Fig.~\ref{fig:B_1}.  Figure~\ref{fig:B_1}
shows that there are no loops present for $\rho<\rho_J$ and therefore the interaction network
must be composed of open paths for such values of $\rho$.   Although the results for the considered $2$D systems and $\rho<\rho_J$ are
similar, we still observe that  $S_{\rm p}$ is consistently
larger for the linear system compared to the non-linear one, in particular for smaller values of $F_{\rm th}$ (compare
Fig.~\ref{fig:shape_selected}(a) and (b)).   

Still considering the $2$D systems, but for $\rho > \rho_{\rm J}$, we observe that for large
$F_{\rm th}$,  $S_{\rm p}$ approaches $1.0$ for both systems, and we can again 
use the observations from Fig.~\ref{fig:B_1}
to conclude that the interaction network is composed of open paths, since
 $B_1/N$ is small.  The differences between linear and non-linear systems appear when smaller
 values of $F_{\rm th}$ are considered: here we find that the values of $S_{\rm p}$ are significantly 
 larger for the linear compared to the non-linear system.  
 The main finding here is that for a significant range of force thresholds around the average force, the 
 clusters are significantly more complex for the systems based on linear compared to nonlinear force model.

Let us now consider the $3$D system, shown in Figs.~\ref{fig:shape}(c) and~\ref{fig:shape_selected}(c).  
First of all, note that 
the data are more noisy in $3$D, particularly for small $F_{\rm th}$. The reason for the noisiness is twofold: 
first, there are only few clusters when $F_{\rm}< 1.0$ at large $\rho$, so the statistics is not very good; second, 
there are also more possible cluster shapes in $3$D compared to $2$D; a quick intuitive explanation comes 
from considering a lattice in $2$D and $3$D and counting the number of possible cluster shapes with a fixed number of particles.
Despite the restrictions imposed by noise, one can still conclude that the clusters are more complex
in $3$D compared to the $2$D (non-linear) system, although the difference between the two is 
not as significant as expected based on geometry considerations (note that 
the upper limit of $S_{\rm p}$  is $7$ and $4$ for $3$D and $2$D geometry, respectively).  
For $F_{\rm th}>2.0$, we find $S_{\rm p}\approx 1.0$. We use again the results shown in 
Fig.~\ref{fig:B_1} to conclude that the interaction network in this regime consists mainly of
open paths.

To summarize, we find that for small or average force thresholds (less than twice the average force) 
regardless of the number of physical dimensions, the non-linear systems contain clusters with less
complex structure compared to the linear system, suggesting that the force model plays a crucial role in the
cluster shape.    
For large force thresholds (considering forces larger than twice the average force), the force networks are typically
composed of open paths for all systems considered, independently of the force model and of the number of
physical dimensions.

\subsection{Persistence diagrams}

In Sec.~\ref{sec:betti} we analyzed $B_0$'s  for a set of force thresholds, $F_{\rm th}$.  Zeroth Betti number 
however provides only the information about the number of clusters, and not about how they are connected.  
This information can be obtained 
by analyzing persistence diagrams, which allow us to capture information about the force clusters for all
force thresholds at once, including their persistence as a considered force threshold is changed. 
We explain the use of the persistence diagrams in interaction network analysis further 
by considering an example below; the reader is referred to~\cite{kramar_2014} for the in-depth discussion of persistent 
homology in the context of interaction networks for particulate systems, and to~\cite{pre13,kramar_pre_14} for less technical descriptions.   
We note that  there are multiple software packages and libraries used to compute persistence diagrams~\cite{perseus,javaPlex,phat,gudhi}. 
We used JavaPlex~\cite{javaPlex} to obtain the results discussed in this section.

When the force threshold, $F_{\rm th}$, changes, the clusters that are formed in the force
network composed of forces above $F_{\rm th}$ either appear or disappear: note that 
a cluster disappears when it merges with another one.  
We indicate a birth, $\theta_{\rm b}$ ($F_{\rm th}$ at which a cluster appears), on
the $x$-axis and death, $\theta_{\rm d}$ ($F_{\rm th}$ when a cluster disappears), on the $y$-axis.
The set of points constructed in this way forms a persistence diagram (PD); we construct a PD for each $\rho$.
Note that for a given $\rho$, a persistence diagram contains the information about particle connectivity for 
all force thresholds at once.  Betti numbers, for example, could be easily computed directly from the persistence
diagrams. However the opposite is not true since the amount of information contained in persistence diagrams is 
significantly larger.  
For the present purposes, we also define the lifespan of a cluster as the difference $\theta_{\rm b}-\theta_{\rm d}$.
The relevance of lifespan in describing the interaction networks is discussed further below.

\begin{figure}[ht!]
  \centering
  \includegraphics[width=3in]{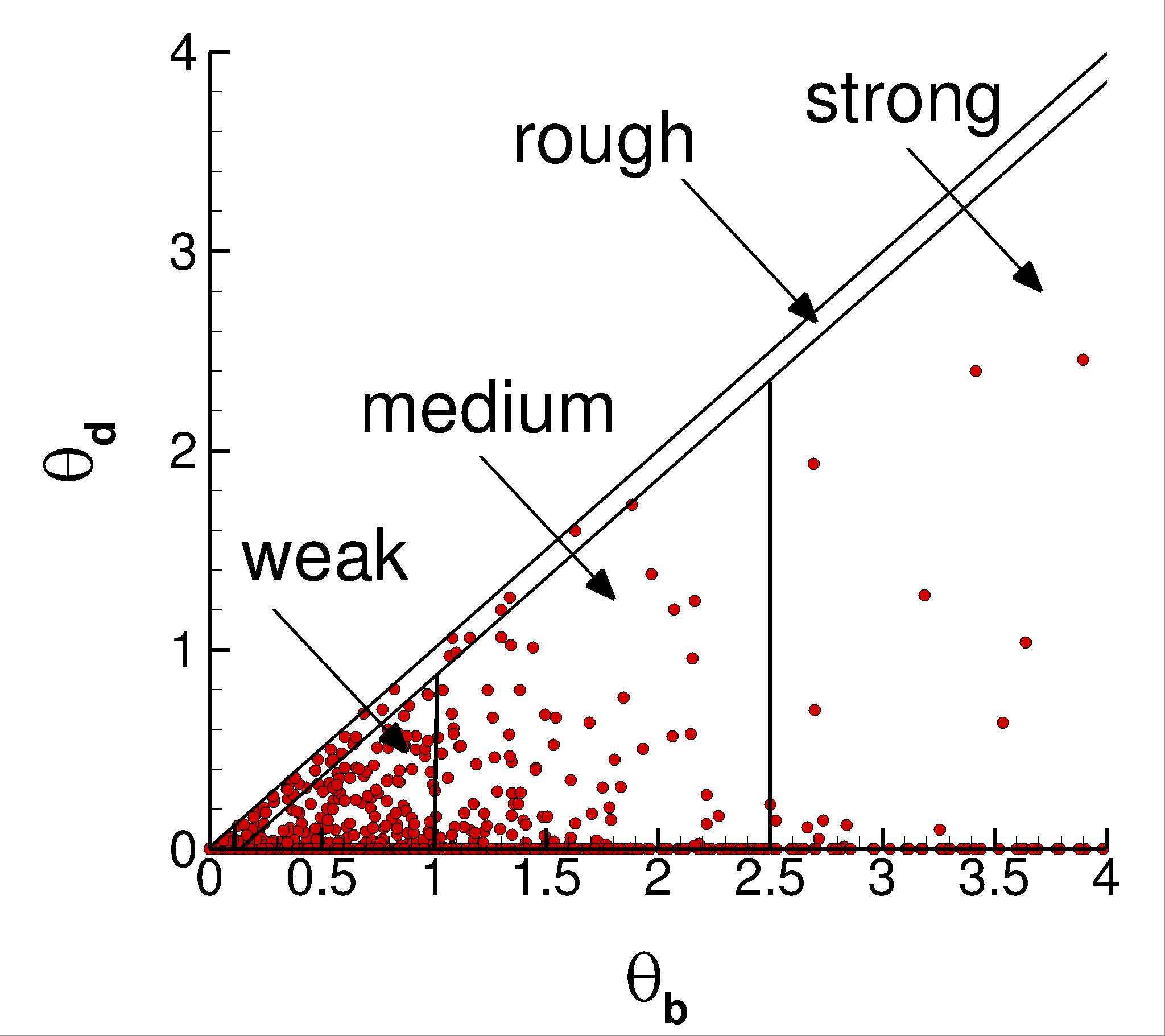}
  \caption{An example of a $B_0$ persistence diagram for $2$D  non-linear force model. 
  This example shows  a separation of the persistence diagram into the following categories: rough, strong, medium and weak
  as discussed in the text.}\label{fig:pers_diagram_example}
\end{figure}

Figure~\ref {fig:pers_diagram_example} shows an example of a PD ($2$D non-linear system is used
here).  One approach to analysis of persistence diagrams is to split them into bins: rough, strong,
medium, and weak, similarly as in~\cite{pre13}. A point on a persistence diagram
belongs to the rough category if $\theta_{\rm b}-\theta_{\rm d}\leq 0.1$.
Such points indicate clusters that have a short lifespan (after the birth, clusters
in the rough region disappear after only a small change in $F_{\rm th}$)
and can be thought of as a noise.
The points that have a lifespan larger than $0.1$ are divided into the following
bins: 
weak when $0.1\leq\theta_{\rm b}<1.0$, 
medium  when $1.0\leq\theta_{\rm b}<2.5$, and 
large when and $2.5\leq\theta_{\rm b}$.   
This binning of the forces is in spirit similar to the separation of interaction networks
in `strong' and `weak' categories, commonly used in the granular literature. We use the binning 
approach for the purpose of developing better understanding of the differences between the 
granular systems considered so far.    Specifically, we will discuss the origins of the $B_0/N$ ridge in Fig.~\ref{fig:B0}.
\begin{figure}[ht!]
  \centering
  \subfloat[]{\includegraphics[width = 1.7in]{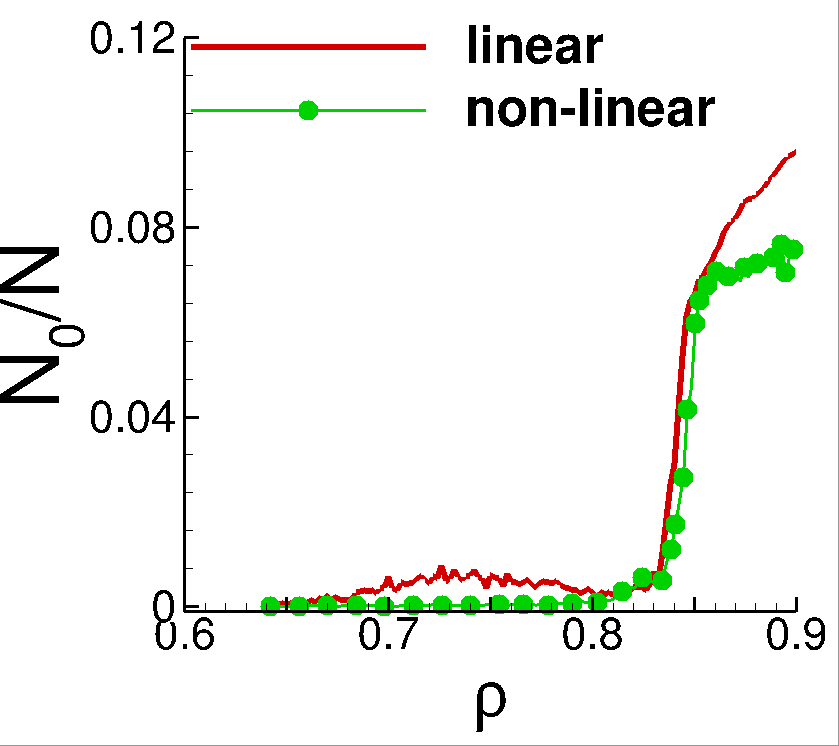}}\subfloat[]{\includegraphics[width = 1.7in]{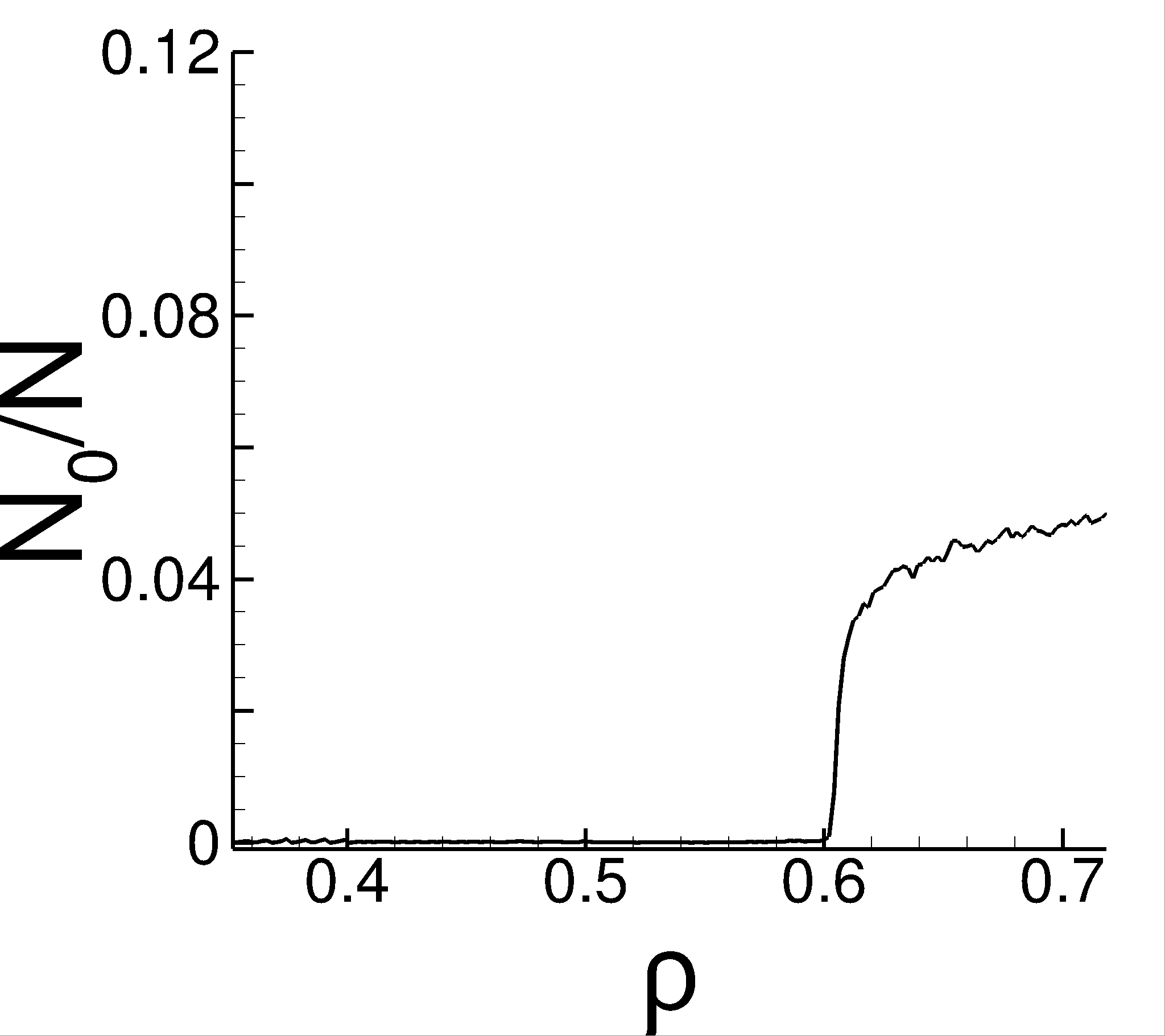}}\\
    \subfloat[]{\includegraphics[width = 1.7in]{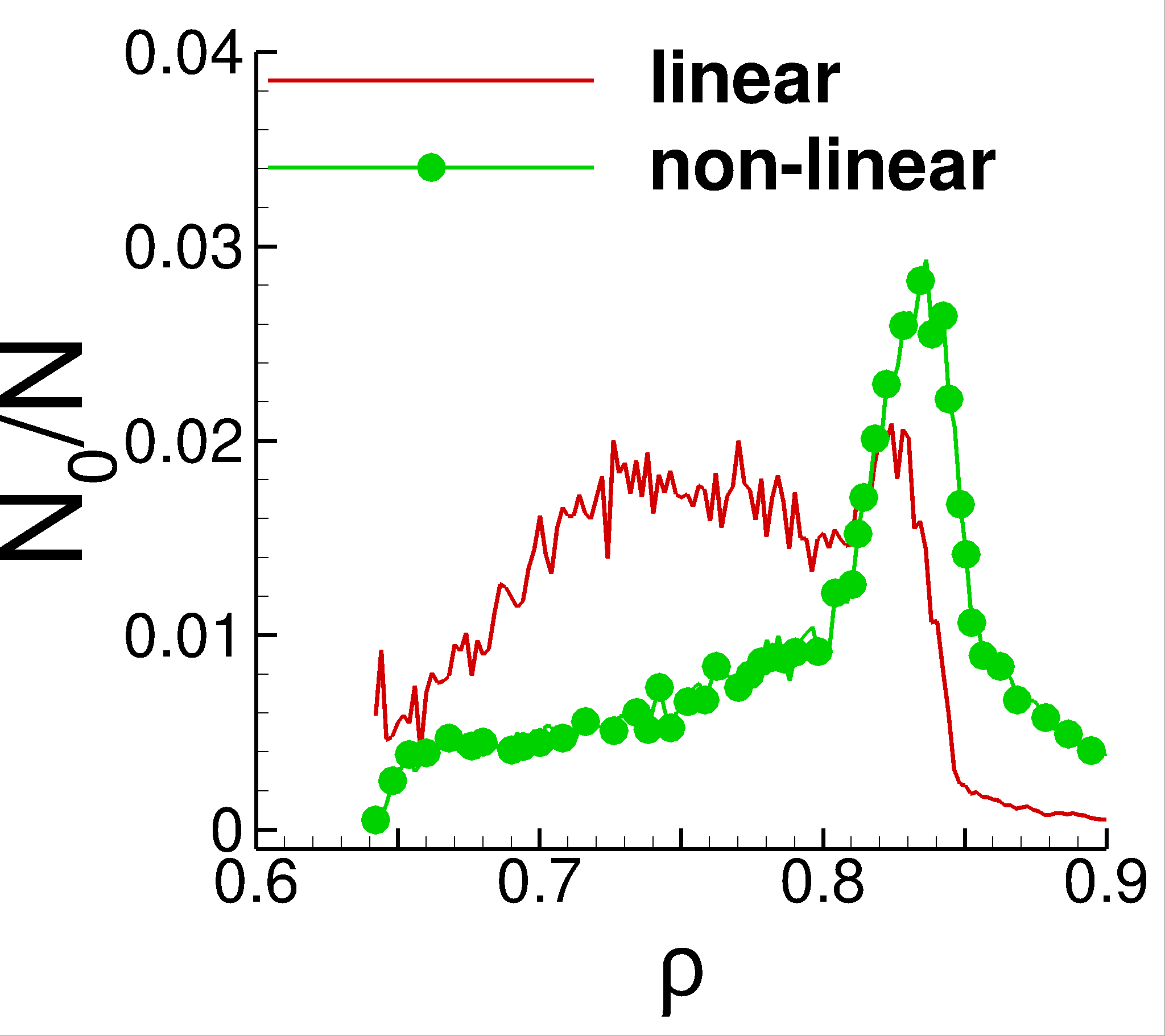}}\subfloat[]{\includegraphics[width = 1.7in]{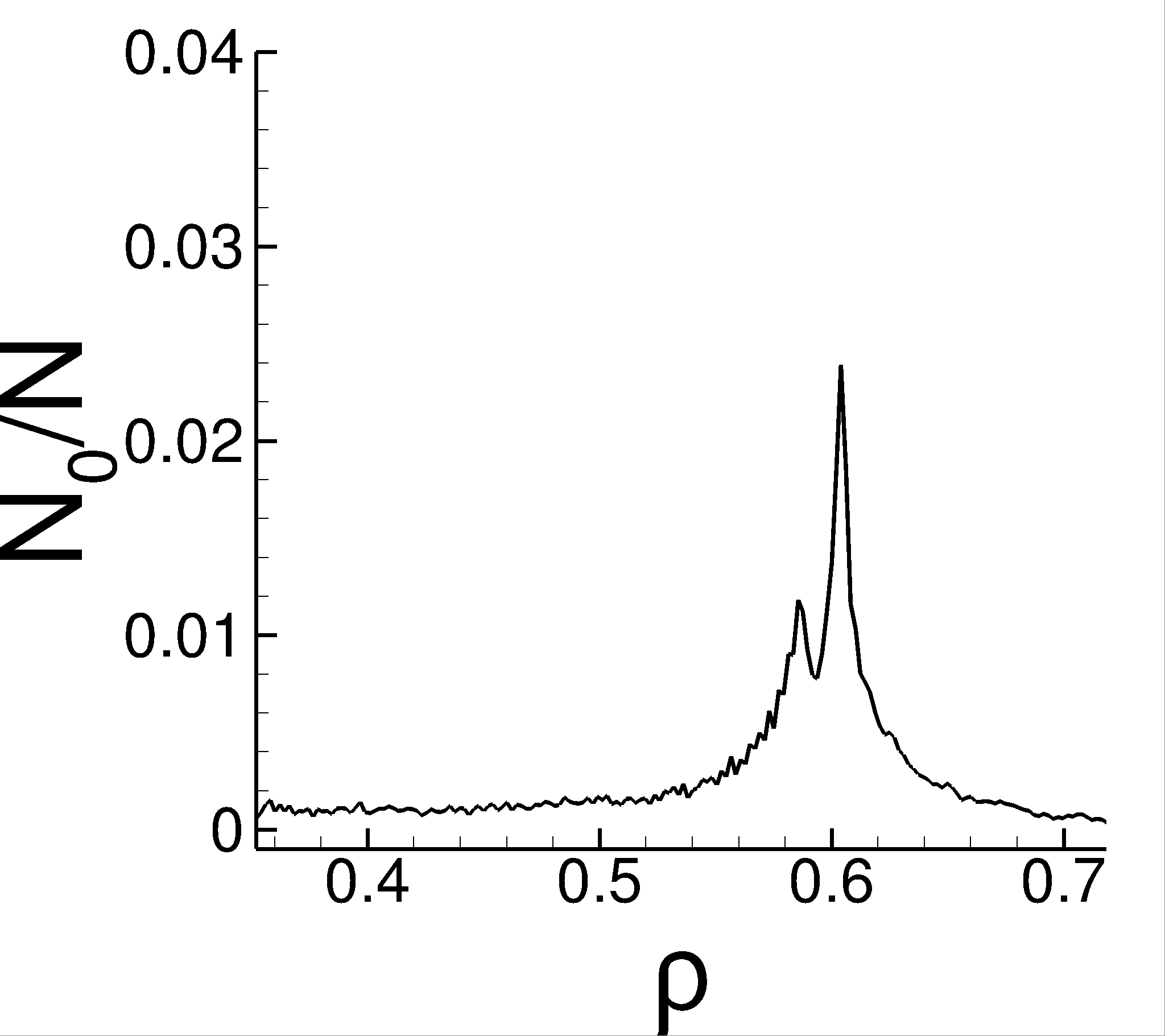}}\\
  \subfloat[]{\includegraphics[width = 1.7in]{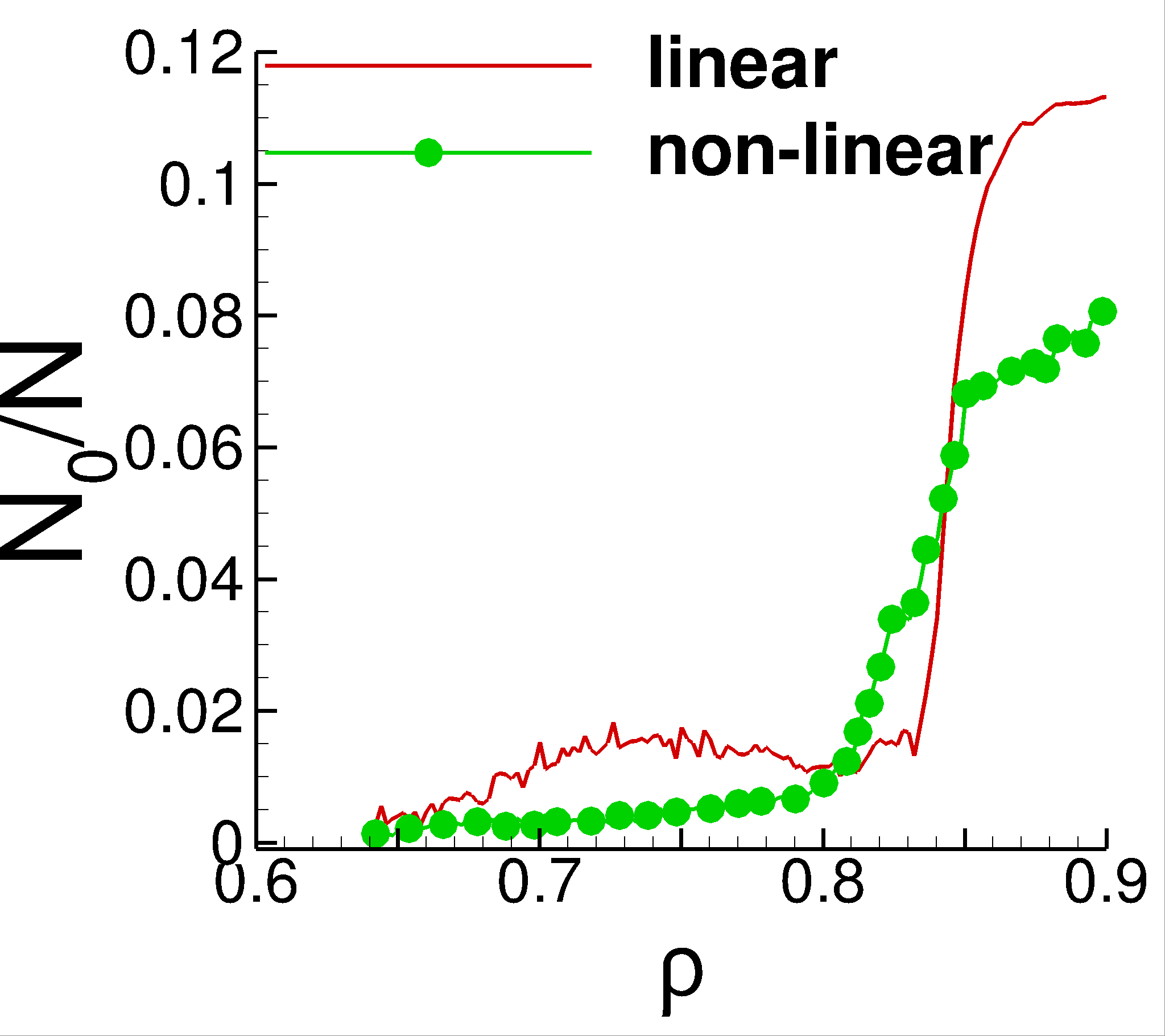}}\subfloat[]{\includegraphics[width = 1.7in]{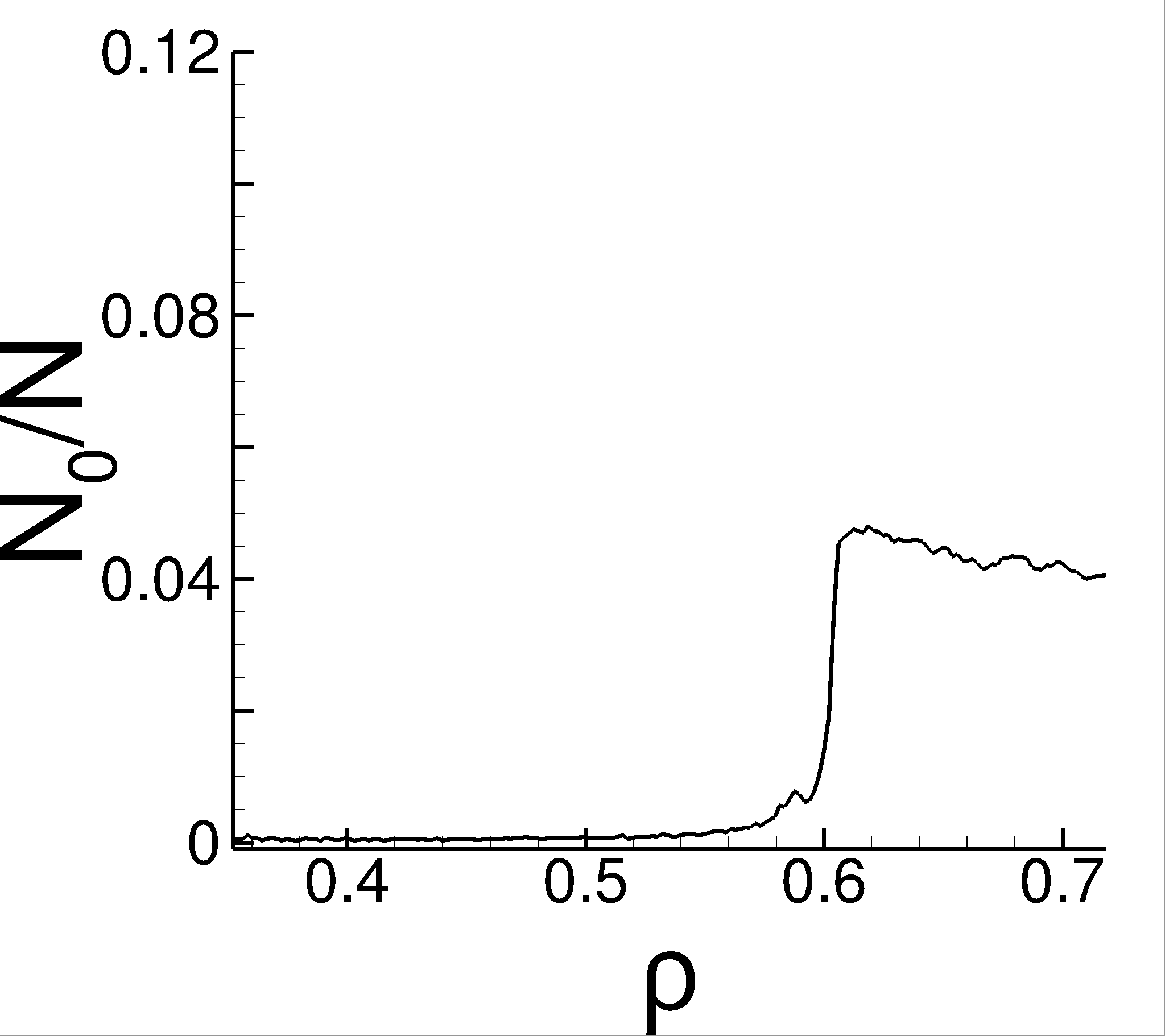}}\\
  \subfloat[]{\includegraphics[width = 1.7in]{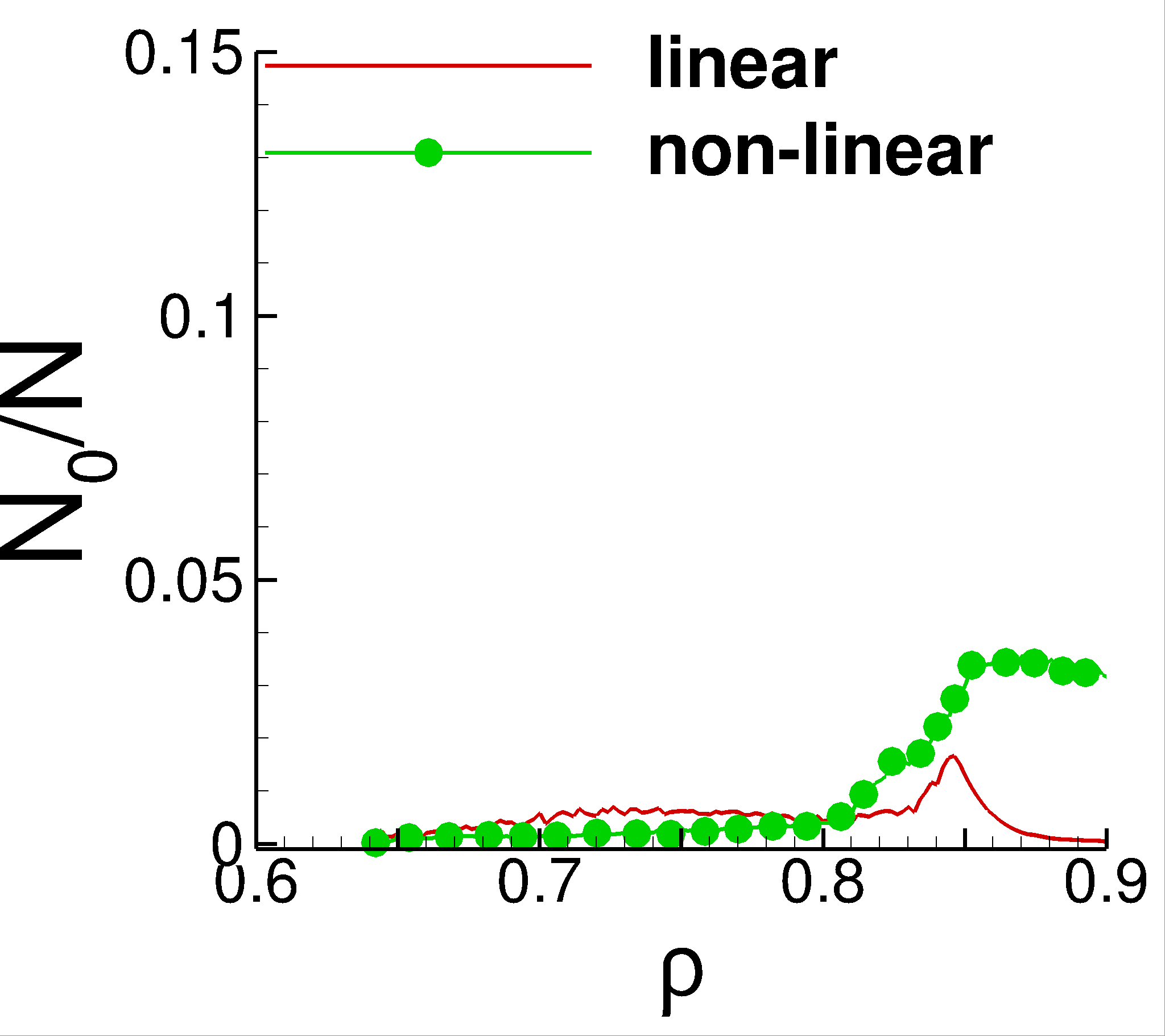}}\subfloat[]{\includegraphics[width = 1.7in]{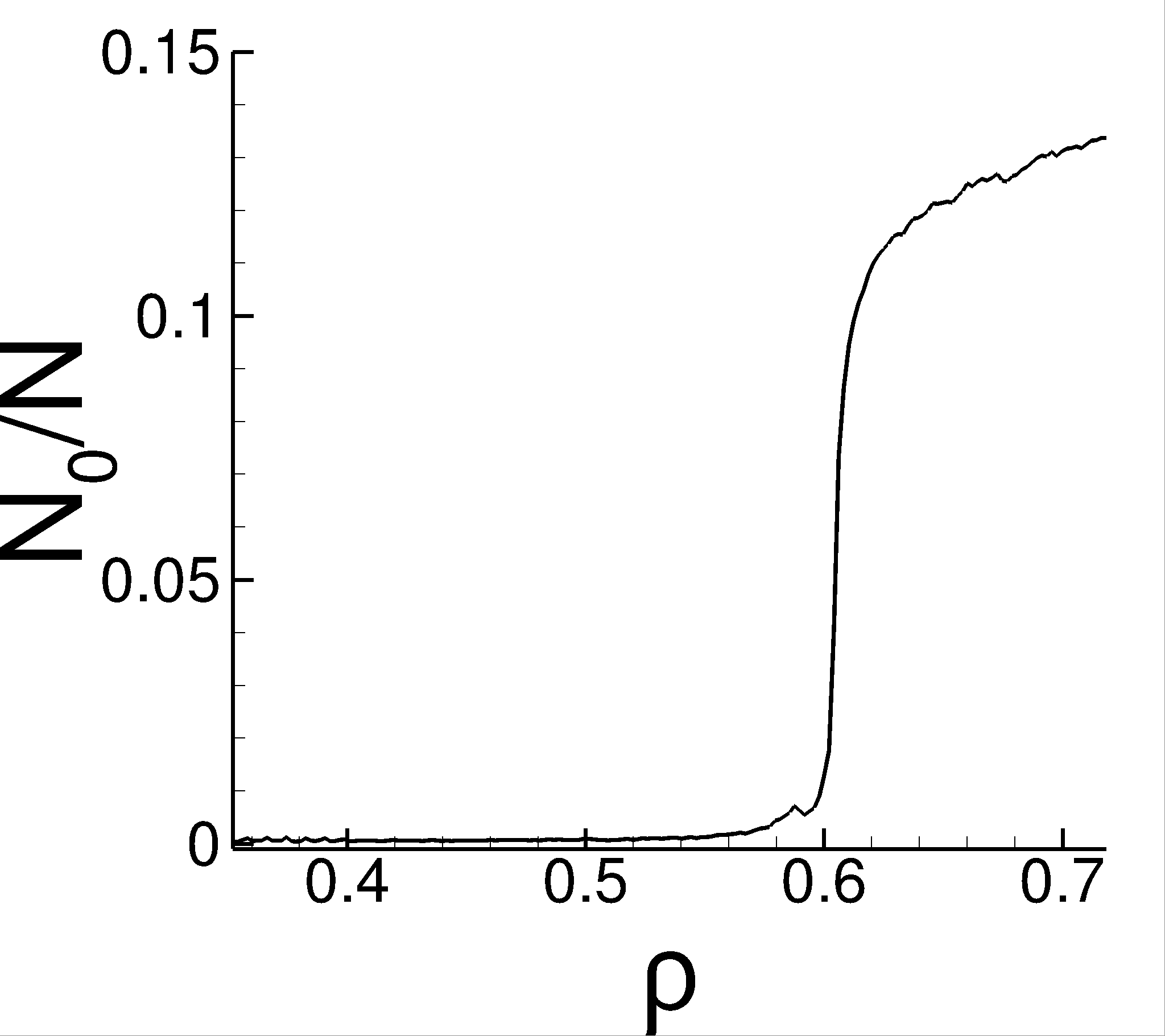}}
  \caption{Average number of points, $N_0$, in different parts of persistence diagrams, normalized by the total number of particles, $N$.
  $1.$ {\it rough}: (a) $2$D and (b) $3$D;
  $2$. {\it  weak}: (c) $2$D and (d) $3$D;
  $3$: {\it medium}: (e) $2$D and (f) $3$D, and
  $4$ {\it strong}: (g) $2$D and (h) $3$D. 
  }\label{fig:persistence}
\end{figure}

Figure~\ref{fig:persistence} shows the average point count in all persistence bins for
the $2$D and $3$D systems; $N_0$ denotes the average number of persistence points
found in the specified category (recall that the results are averaged over $20$ realizations),
and $N$ is the total number of particles.

To start the discussion of Fig.~\ref{fig:persistence}, we first comment on the results relevant to 
the bins for which the differences between considered systems are minor: rough and medium.
Regarding the rough regime, Fig.~\ref{fig:persistence}(a) - (b), we observe that  the roughness of the force
network is very similar near jamming for all the systems.  For the medium bin, see 
Fig.~\ref{fig:persistence}(e) - (f), we observe a sudden increase of the curves beyond jamming.   
For both rough and medium bins, we note a smaller number of points for the considered 
non-linear systems; we comment about this result in the case of medium bins further below. 

The largest differences between the systems based on different force models can be observed
for strong, Fig.~\ref{fig:persistence}(g) - (h) and weak, Fig.~\ref{fig:persistence}(c) - (d) bins. 
Considering first the strong bin, we note that the $2$D system based on the linear force model 
develops a peak of $N_0/N$ during jamming, while for the non-linear systems such a peak is 
lacking, since these systems show large values of $N_0/N$ beyond jamming as well. Such a peak
during jamming can be explained by strong collisions of the particles that form relevant clusters.
For the system based on the linear force model, for $\rho> \rho_{\rm J}$, the majority of the clusters 
remain in the medium bin, see Fig.~\ref{fig:persistence}(e) (note that a similar behavior was 
observed for the $2$D systems in continuously compressed systems~\cite{pre13}).
However, for non-linear force model, the clusters formed by strong collisions remain
in the strong bin as $\rho$ increases.   Clearly, the softness of the non-linear interaction potential 
plays an important role and influences strongly the properties of the interaction networks as a system is 
compressed beyond jamming.   One consequence of this difference between linear and non-linear
systems is larger spread of forces for the non-linear systems for $\rho> \rho_{\rm J}$.  Consistently,  
as already observed in Fig.~\ref{fig:PDF}, for $\rho> \rho_{\rm J}$, the PDF's of the non-linear systems are 
wider compared to the linear case.

In contrast, Fig.~\ref{fig:persistence}(c) - (d) shows a pronounced peak in the weak force bin
around jamming for the systems based on the non-linear interaction model, but not for the linear one. 
Recalling formation of the ridge in $B_0$ (viz. Fig.~\ref{fig:B_0})  for similar packing fractions, 
we conjecture that the ridge is formed by weak interaction that dominate the interaction network 
for $\rho \approx \rho_{\rm J}$.  

Another measure based on the persistence diagrams is total persistence~\cite{pre13}
of a PD, defined as
\begin{equation*}
 {\rm TP}(\rm{PD}) = \sum_{\rm (\theta_{\rm b},\theta_{\rm d})\in PD}{(\theta_{\rm b}-\theta_{\rm d})}
\end{equation*}
where the sum ranges over all $(\theta_{\rm b},\theta_{\rm d})$ pairs corresponding to the points in 
the $B_{\rm 0}$ and $B_{\rm 1}$ persistence diagrams.  For simplicity of notation, we refer to the 
total persistence for the clusters by TP$_{\rm 0}$, and for the loops by TP$_{\rm 1}$. 
The physical interpretation of these quantities could be best described by a landscape analogy: 
if we think of an interaction network as a landscape, then large TP$_{\rm 0}$ implies a landscape containing
a large number of prominent peaks and valleys, and large TP$_{\rm 1}$ suggests well developed
connectivity between the peaks (leading to loops).    Note that the concept of force threshold is 
not relevant anymore here, since total persistence includes the information about all force thresholds 
at once. 

Next, we discuss briefly TP$_{\rm 0}$ and TP$_{\rm 1}$ for the considered systems. 
A detailed discussion regarding TP$_{\rm 2}$ is omitted since
this measure is defined only in $3$D. It suffices to  mention that TP$_{\rm 2} \approx 0$ for
$\rho<\rho_J$ and increases monotonously for $\rho> \rho_{\rm J}$.  

\begin{figure}[ht!]
  \centering
  \subfloat[]{\includegraphics[width=1.7in]{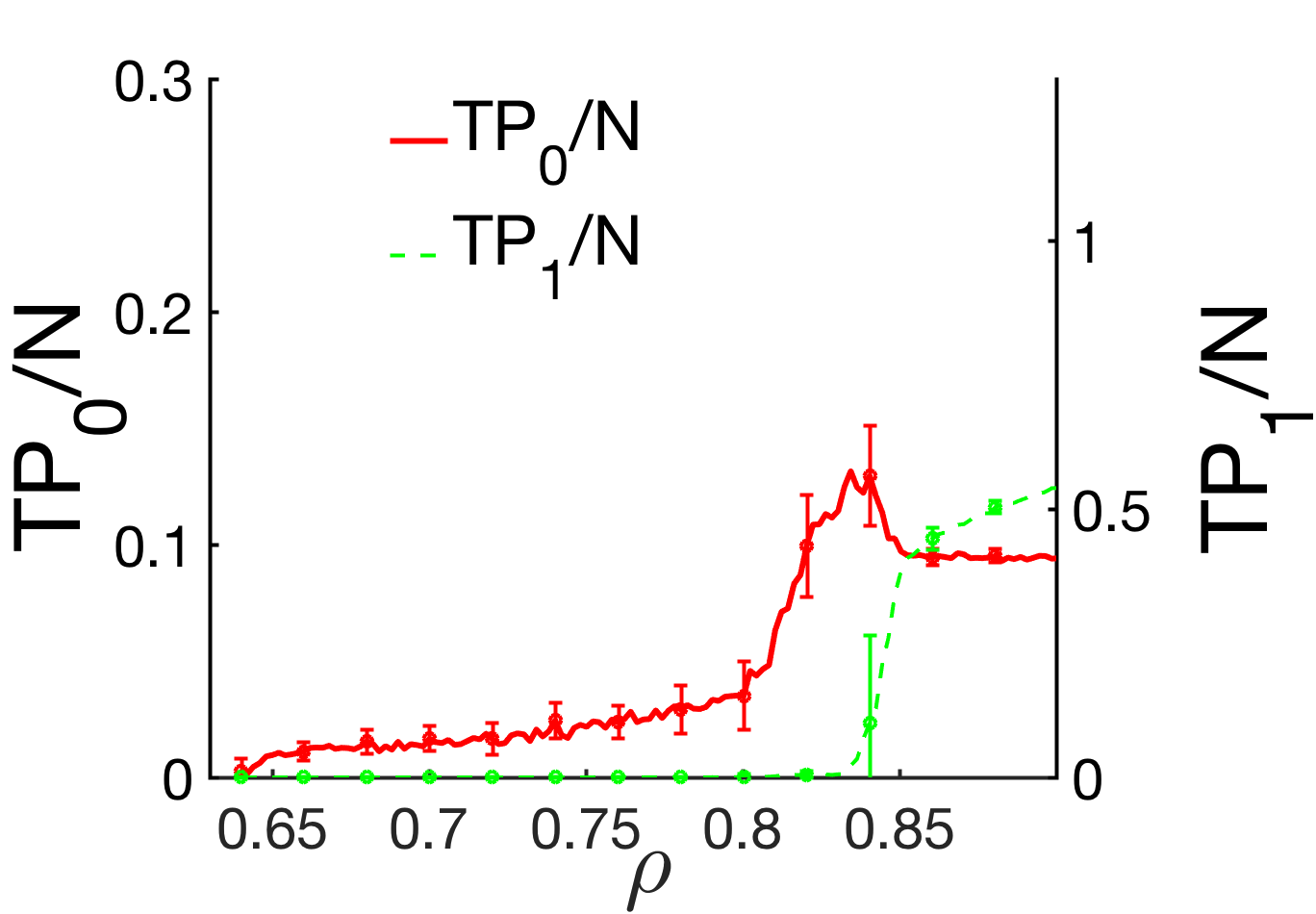}}
  \subfloat[]{\includegraphics[width=1.7in]{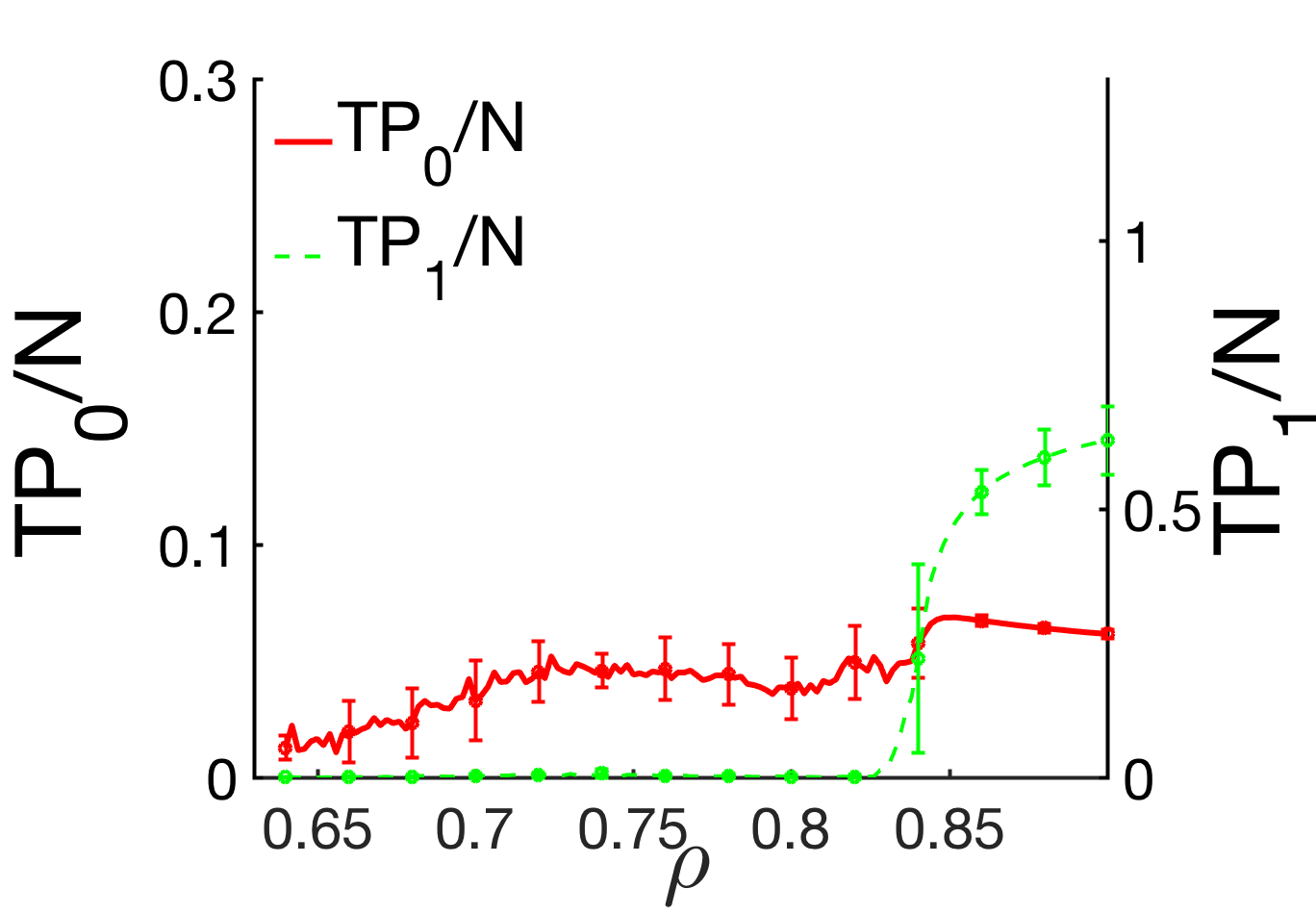}}\\
  \subfloat[]{\includegraphics[width=1.7in]{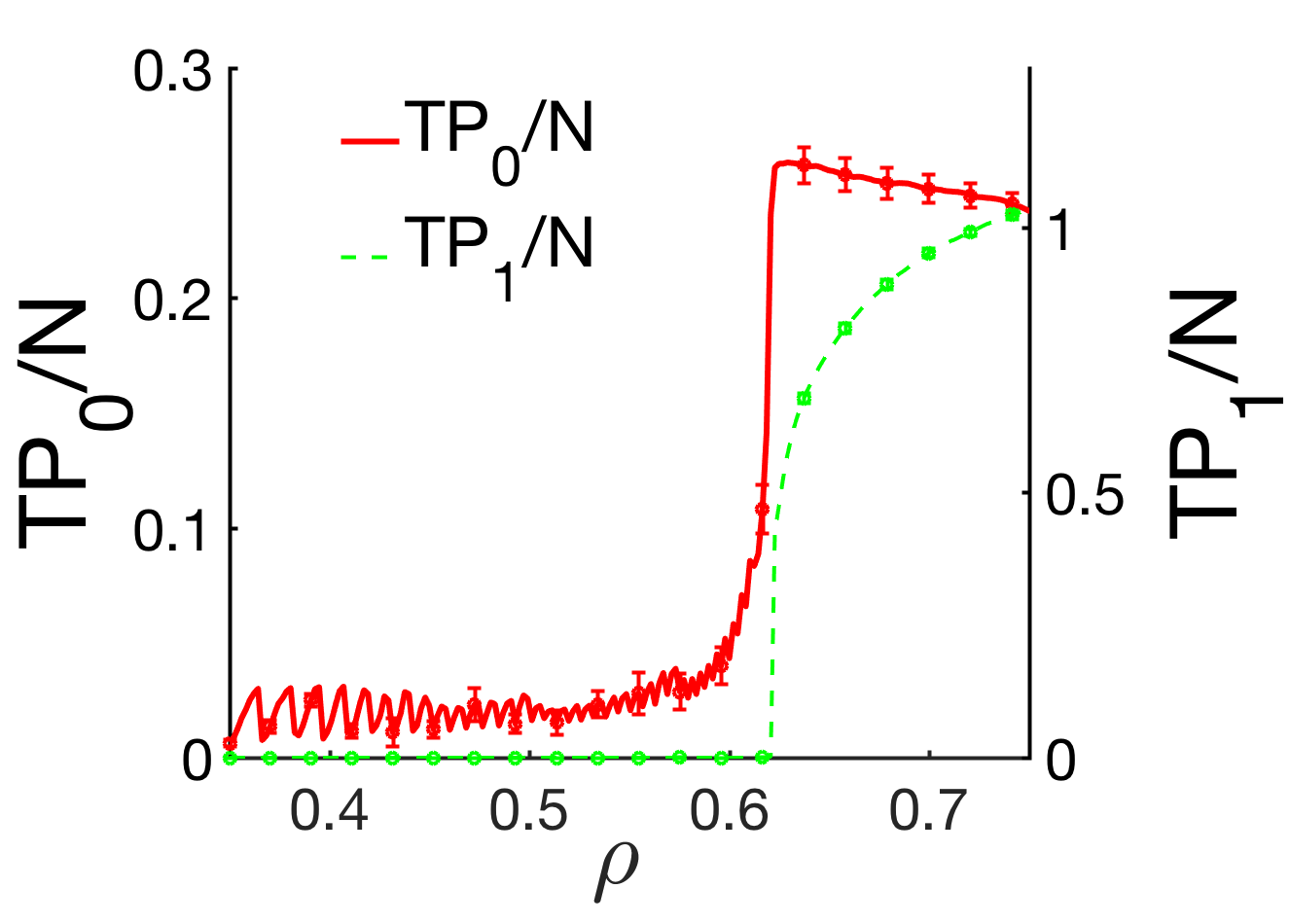}}
  \caption{Total persistence in the 2D systems
  for (a) $2$D non-linear force model,  (b) 2D linear force model and (c) $3$D system;
 $N$ is the total number of particles.
  }\label{fig:TP}
\end{figure} 

Figure~\ref{fig:TP} shows TP$_{\rm 0}$ and TP$_{\rm 1}$ for the considered $2$D and $3$D systems.  
The most obvious finding that is relevant to both TP$_{\rm 0}$ and TP$_{\rm 1}$ is 
a significantly more prominent increase of these measures close to $\rho_{\rm J}$ for the $3$D system, 
compared to the $2$D ones.   This result suggests that additional physical dimension leads to an increasingly 
complex force landscape that includes a larger number of prominent peaks and valleys.  To understand this landscape more precisely, 
one needs to consider in more detail the relevant persistence diagrams.  Such more in-depth analysis is left 
for future work. 

To reiterate the findings discussed in this section, we find that both the number of physical dimensions 
and the force model play a role in determining the topology of interaction networks.
The number of physical dimensions influences the behavior of  the percolating cluster in the
interaction network, and the $B_0$ results show that the force model is important when we consider
small forces in our computations of interaction network properties.   We also find that the properties of the
interaction networks during jamming transition can be quantified by persistence 
diagrams and derived quantities, such as total persistence.   

\section{Conclusions}
\label{sec:conclusions}

In this work, we analyze the properties of interaction networks with a focus on
the influence of the number of physical dimensions and of the force model (linear vs. non-linear) describing 
particle interactions.
The comparison of the force distribution reveals differences between the considered systems for a large 
range of packing fractions, $\rho$.  
Specifically,  the
force distributions are found to be very different between $2$D and $3$D systems in the case of small $\rho$'s, 
despite consistent preparation and relaxation protocols.  
Beyond jamming, we find that the distribution of forces is wider for the
non-linear systems regardless of the number of physical dimensions. 
This being said, we emphasize that our results were obtained mainly for one system size; while
the general features of the results were found to hold as the system size was varied, no systematic analysis of
the influence of system size on our results has been carried out.

The analysis of the percolating cluster and its size as the force threshold, $F_{\rm th}$, is varied  for $\rho\geq\rho_J$
shows that a percolating cluster is composed of a significant number of particles in $2$D (often at least the half of 
the total number of particles) at any time percolation occurs. On the other hand, in $3$D, we find percolating clusters
composed of a very small number of particles for the granular 
system close to jamming or when $F_{\rm th}$ is large, typically for $F_{\rm th}>2.0$.   For small force 
thresholds, we offer an explanation of this finding based solely on the number of physical dimensions.

Our force distribution and percolation analysis results suggest important structural and 
topological differences between the $2$D and $3$D systems, as well as between the systems 
that are based on different force models. This motivates
the analysis of the average cluster size, $\langle S\rangle/N$, shape, $S_p$,
and the first two Betti numbers, $B_0$ and $B_1$.

The average cluster size, $\langle S\rangle/N$, in an interaction network (not including the percolating cluster) shows
a peak formation for small $F_{\rm th}\approx 0$ close to jamming transition.  Such a peak 
occurs only for the granular systems based on non-linear interactions. The differences between 
$2$D and $3$D systems arise beyond jamming; for $3$D there are larger clusters characterized by a 
larger interaction force (when normalized by the average force) compared to $2$D.  

One of the prominent results of this study is the formation of a pronounced ridge in the number of 
clusters at $\rho \approx \rho_{\rm J }$ for the non-linear force models.   The formation and the properties of 
this ridge, that becomes visible when considering weak interaction network, have been 
carefully analyzed using topological measures.  Our conjecture is that the softness of non-linear interaction 
model plays a significant role in determining the properties of the interaction networks as the systems go through 
jamming.   The consequence is that close to jamming, one could expect significantly different behavior of the systems involving 
particles interacting by different interaction laws.
It should be pointed out that most of our conclusions were obtained by considering systems of a given 
size.  While the main features of the results were found to be robust as the system size is moderately varied, 
detailed analysis of the influence of system size on the results remains to be carried out.

In conclusion, based on a well-defined set of measures, in this work we provide a precise and objective comparison
of the interaction networks in finite size compressed granular systems.  The main finding is that both the nature of the 
interactions between the constitutive particles, and the number of physical dimensions, play a significant
role in determining the interaction networks' properties.  In contrast, the interaction networks are 
found to be only weakly influenced by friction between the particles.   
It remains to be seen to which degree
these findings extend to more complex systems exposed to shear or other types of external 
influences.
\vspace{0.1in}
\begin{acknowledgments}
The initial stages of this work were carried out as a part of undergraduate 
Capstone class at NJIT~\cite{capstone}, and we thank to all the participating students for their 
help and contributions towards understanding of interaction networks in granular systems. 
This research was supported in 
part by  the NSF Grant No. DMS-1521717, DARPA contract No. HR0011-16-2-0033, 
and ARO Grant No. W911NF1810184.

\end{acknowledgments}


\begin{thebibliography}{10}

\bibitem{arevalo_bif_chaos_09}
R.~Ar\'evalo, I.~Zuriguel, and D.~Maza.
\newblock Topological properties of the contact network of granular materials.
\newblock {\em Int. J. Bif. Chaos}, 19(02):695--702, 2009.

\bibitem{arevalo_pre10}
R.~Ar\'evalo, I.~Zuriguel, and D.~Maza.
\newblock Topology of the force network in jamming transition of an
  isotropically compressed granular packing.
\newblock {\em Phys. Rev. E}, 81:041302, 2010.

\bibitem{arevalo_pre13}
Roberto Ar\'evalo, Luis~A. Pugnaloni, Iker Zuriguel, and Diego Maza.
\newblock Contact network topology in tapped granular media.
\newblock {\em Phys. Rev. E}, 87:022203, Feb 2013.

\bibitem{azema_pre_12}
E.~Az\'ema and F.~Radja\"{\i}.
\newblock Force chains and contact network topology in sheared packings of
  elongated particles.
\newblock {\em Phys. Rev. E}, 85:031303, 2012.

\bibitem{ardanza_pre14}
S.~Ardanza-Trevijano, Iker Zuriguel, Roberto Ar\'evalo, and Diego Maza.
\newblock Topological analysis of tapped granular media using persistent
  homology.
\newblock {\em Phys. Rev. E}, 89:052212, May 2014.

\bibitem{epl12}
L.~Kondic, A.~Goullet, C.S. O'Hern, M.~Kramar, K.~Mischaikow, and R.P.
  Behringer.
\newblock Topology of force networks in compressed granular media.
\newblock {\em Europhys. Lett.}, 97:54001, 2012.

\bibitem{pre13}
M.~Kram\'ar, A.~Goullet, L.~Kondic, and K.~Mischaikow.
\newblock Persistence properties of compressed granular matter.
\newblock {\em Phys. Rev. E}, 87:042207, 2013.

\bibitem{kramar_2014}
M.~Kram\'ar, A.~Goullet, L.~Kondic, and K.~Mischaikow.
\newblock Quantifying force networks in particulate systems.
\newblock {\em Physca D.}, 283:37 -- 55, 2014.

\bibitem{kramar_pre_14}
M.~Kram\'ar, A.~Goullet, L.~Kondic, and K.~Mischaikow.
\newblock Evolution of force networks in dense particulate media.
\newblock {\em Phys. Rev. E}, 90:052203, 2014.

\bibitem{Pugnaloni_2016}
L.A. Pugnaloni, C.M. Carlevaro, M.~Kram\'ar, K.~Mischaikow, and L.~Kondic.
\newblock Structure of force networks in tapped particulate systems of disks
  and pentagons. i. clusters and loops.
\newblock {\em Phys. Rev. E}, 93:062902, 2016.

\bibitem{kondic_2016}
L.~Kondic, M.~Kram\'ar, Luis~A. Pugnaloni, C.~Manuel Carlevaro, and
  K.~Mischaikow.
\newblock Structure of force networks in tapped particulate systems of disks
  and pentagons. ii. persistence analysis.
\newblock {\em Phys. Rev. E}, 93:062903, Jun 2016.

\bibitem{pg17}
L.~Kondic, M.~Kramar, L.~Kovalcinova, and K.~Mischaikow.
\newblock Evolution of force networks in dense granular matter close to
  jamming.
\newblock {\em EPJ Web of Conferences}, 140:15014, 2017.

\bibitem{PRE_2018}
J.~A. Dijksman, L.~Kovalcinova, J.~Ren, R.~P. Behringer, M.~Kram\'ar,
  K.~Mischaikow, and L.~Kondic.
\newblock Characterizing granular networks using topological metrics.
\newblock {\em Phys. Rev. E}, 97:042903, 2018.

\bibitem{pre18_impact}
T.~Takahashi, Abram~H. Clark, T.~Majmudar, and L.~Kondic.
\newblock Granular response to impact: Topology of the force networks.
\newblock {\em Phys. Rev. E}, 97:012906, 2018.

\bibitem{zhang_10}
J.~Zhang, T.~S. Majmudar, A.~Tordesillas, and R.~P. Behringer.
\newblock Statistical properties of a 2d granular material subjected to cyclic
  shear.
\newblock {\em Granul. Matter}, 12:159--172, Apr 2010.

\bibitem{tordesillas_pre10}
A.~Tordesillas, D.~M. Walker, and Q.~Lin.
\newblock Force cycles and force chains.
\newblock {\em Phys. Rev. E}, 81:011302, 2010.

\bibitem{walker_pre12}
D.M. Walker and A.~Tordesillas.
\newblock Taxonomy of granular rheology from grain property networks.
\newblock {\em Phys. Rev. E}, 85:011304, 2012.

\bibitem{peters05}
J.P. Peters, M.~Muthuswamy, J.~Wibowo, and A.~Tordesillas.
\newblock Characterization of force chains in granular material.
\newblock {\em Phys. Rev. E}, 72:041307, 2005.

\bibitem{daniels_pre12}
Danielle~S. Bassett, Eli~T. Owens, Karen~E. Daniels, and Mason~A. Porter.
\newblock Influence of network topology on sound propagation in granular
  materials.
\newblock {\em Phys. Rev. E}, 86:041306, 2012.

\bibitem{tighe_sm10}
Brian~P. Tighe, Jacco~H. Snoeijer, Thijs J.~H. Vlugt, and Martin van Hecke.
\newblock The force network ensemble for granular packings.
\newblock {\em Soft Matter}, 6:2908--2917, 2010.

\bibitem{tighe_jsm11}
B.~P. Tighe and J.H.T. Vlugt.
\newblock Stress fluctuations in granular foce networks.
\newblock {\em J. Stat. Mech.}, 11:04002, 2011.

\bibitem{sarkar_prl13}
Sumantra Sarkar, Dapeng Bi, Jie Zhang, R.~P. Behringer, and Bulbul Chakraborty.
\newblock Origin of rigidity in dry granular solids.
\newblock {\em Phys. Rev. Lett.}, 111:068301, Aug 2013.

\bibitem{radjai_96b}
F.~Radjai, M.~Jean, J.~J. Moreau, and S.~Roux.
\newblock Force distribution in dense two-dimensional granular systems.
\newblock {\em Phys. Rev. Lett.}, 77:274--277, 1996.

\bibitem{radjai97b}
F.~Radjai, D.~E. Wolf, S.~Roux, M.~Jean, and J.~J. Moreau.
\newblock Force networks in dense granular media.
\newblock In R.~P. Behringer and J.~T. Jenkins, editors, {\em Powders \& Grains
  97}, pages 211--214. Balkema, Rotterdam, 1997.

\bibitem{radjai99}
F.~Radjai, J.~J. Moreau, and S.~Roux.
\newblock Contact forces in a granular packing.
\newblock {\em Chaos}, 9:544--550, 1999.

\bibitem{ostojic06}
S.~Ostojic, E.~Somfai, and B.~Nienhuis.
\newblock Scale invariance and universality of force networks in static
  granular matter.
\newblock {\em Nature}, 439:828--830, 2006.

\bibitem{kovalcinova_scaling}
L.~Kovalcinova, A.~Goullet, and L.~Kondic.
\newblock Scaling properties of force networks for compressed particulate
  systems.
\newblock {\em Phys. Rev. E}, 93:042903, 2016.

\bibitem{ostojic_pre07}
S.~Ostojic, T.J.H. Vlugt, and B.~Nienhuis.
\newblock Universal anisotropy in force networks under shear.
\newblock {\em Phys. Rev. E}, 75(3):030301, 2007.

\bibitem{stauffer}
D.~Stauffer and A.~Aharonov.
\newblock {\em Introduction to {P}ercolation {T}heory}.
\newblock Taylor \& Francis, 2003.

\bibitem{kovalcinova_percolation}
L.~Kovalcinova, A.~Goullet, and L.~Kondic.
\newblock Percolation and jamming transitions in particulate systems with and
  without cohesion.
\newblock {\em Phys. Rev. E}, 92:032204, 2015.

\bibitem{ohern01}
C.S. O'Hern, S.A. Langer, A.J. Liu, and S.R. Nagel.
\newblock Force distributions near jamming and glass transitions.
\newblock {\em Phys. Rev. Lett.}, 86:111--114, 2001.

\bibitem{ohern_pre03}
C.~S. O'Hern, L.~E. Silbert, A.~J. Liu, and S.~A. Langer.
\newblock Jamming at zero temperature and zero applied stress: the epitome of
  disorder.
\newblock {\em Phys. Rev. E}, 68:011306, 2003.

\bibitem{ohern_pre_12}
T.~Shen, Corey~S. O'Hern, and M.~D. Shattuck.
\newblock Contact percolation transition in athermal particulate systems.
\newblock {\em Phys. Rev. E}, 85:011308--4, 2012.

\bibitem{Pastor-Satorras_12}
R.~Pastor-Satorras and M.-C. Miguel.
\newblock Percolation analysis of force networks in anisotropic granular
  matter.
\newblock {\em J. Stat. Mech.}, 2012(2), 2012.

\bibitem{majmudar05a}
T.~S. Majmudar and R.~P. Behringer.
\newblock Contact force measurements and stress-induced anisotropy in granular
  materials.
\newblock {\em Nature}, 435:1079--1082, 2005.

\bibitem{clark_prl12}
Abram~H. Clark, Lou Kondic, and Robert~P. Behringer.
\newblock Particle scale dynamics in granular impact.
\newblock {\em Phys. Rev. Lett.}, 109:238302, Dec 2012.

\bibitem{brodu_natcom_15}
N.~Brodu, J.~Dijksman, and R.P. Behringer.
\newblock Spanning the scales of granular materials through microscopic force
  imaging.
\newblock {\em Nat. Commun.}, 6:6361, 2015.

\bibitem{Saadatfar_2012}
M.~Saadatfar, A.~P. Sheppard, T.~J. Senden, and A.~J. Kabla.
\newblock Mapping forces in a 3d elastic assembly of grains.
\newblock {\em J. Mech. Phys. Solids}, 60:55 -- 66, 2012.

\bibitem{brujic_2003}
J.~{Bruji{\'c}}, S.~F. {Edwards}, D.~V. {Grinev}, I.~{Hopkinson},
  D.~{Bruji{\'c}}, and H.~A. {Makse}.
\newblock {3D bulk measurements of the force distribution in a compressed
  emulsion system}.
\newblock {\em Faraday Discuss.}, 123:207--220, 2003.

\bibitem{hurley_prl16}
R.~C. Hurley, S.~A. Hall, J.~E. Andrade, and J.~Wright.
\newblock Quantifying interparticle forces and heterogeneity in 3d granular
  materials.
\newblock {\em Phys. Rev. Lett.}, 117:098005, Aug 2016.

\bibitem{silbert02a}
Leonardo~E. Silbert, Gary~S. Grest, and James~W. Landry.
\newblock Statistics of the contact network in frictional and frictionless
  granular packings.
\newblock {\em Phys. Rev. E}, 66:061303, 2002.

\bibitem{Papadopoulos_18}
L.~Papadopoulos, M.~A Porter, K.~E. Daniels, and D.~S. Bassett.
\newblock Network analysis of particles and grains.
\newblock {\em J Complex Networks}, page 005, 2018.

\bibitem{supp_mat}
Animation of the shear cycle.

\bibitem{radjai97}
F.~Radjai, D.~Wolf, S.~Roux, M.~Jean, and J.~J. Moreau.
\newblock Force networks in granular packings.
\newblock In D.~E. Wolf and P.~Grassberger, editors, {\em Friction, Arching and
  Contact Dynamics}, Singapore, 1997. World Scientific.

\bibitem{perseus}
V.~Nanda.
\newblock Perseus, 2012.

\bibitem{javaPlex}
H.~Adams, A.~Tausz, and M.~Vejdemo-Johansson.
\newblock javaplex: A research software package for persistent (co)homology.
\newblock In Hoon Hong and Chee Yap, editors, {\em Mathematical Software --
  ICMS 2014}, pages 129--136, Berlin, Heidelberg, 2014. Springer Berlin
  Heidelberg.

\bibitem{phat}
Ulrich Bauer, Michael Kerber, Jan Reininghaus, and Hubert Wagner.
\newblock {PHAT}: Persistent homology algorithms toolbox.
\newblock {\em J. Symbolic Computation}, 78:76 -- 90, 2017.

\bibitem{gudhi}
{GUDHI}: Geometry understanding in higher dimensions.
\newblock \url{http://gudhi.gforge.inria.fr}.

\bibitem{ostojic}
S.~Ostojic, E.~Somfai, and B.~Nienhuis.
\newblock Scale invariance and universality of force networks in static
  granular matter.
\newblock {\em Nature}, 439(7078):828--830, Feb 16 2006.

\bibitem{kondic_99}
L.~Kondic.
\newblock Dynamics of spherical particles on a surface: Collision-induced
  sliding and other effects.
\newblock {\em Phys. Rev. E}, 60:751--770, 1999.

\bibitem{brodu_pre_15}
N.~Brodu, J.~A. Dijksman, and R.~P. Behringer.
\newblock Multiple-contact discrete-element model for simulating dense granular
  media.
\newblock {\em Phys. Rev. E}, 91:032201, 2015.

\bibitem{Note1}
Minor influence of different material properties of the particles could be due
  to particle - wall interaction, and, in the case of a nonlinear force model,
  due to weak dependence of the particle collision time on impact speed.

\bibitem{cundall79}
P.~A. Cundall and O.~D.~L. Strack.
\newblock A discrete numerical model for granular assemblies.
\newblock {\em G\'eotechnique}, 29:47--65, 1979.

\bibitem{brendel98}
L.~Brendel and S.~Dippel.
\newblock Lasting contacts in molecular dynamics simulations.
\newblock In {\em Physics of Dry Granular Media}, page 313, Dordrecht, 1998.
  Kluwer Academic Publishers.

\bibitem{goldhirsch_nature05}
C.~Goldenberg and I.~Goldhirsch.
\newblock Friction enhances elasticity in granular solids.
\newblock {\em Nature}, 435:188, 2005.

\bibitem{luding94d}
S.~Luding, E.~Cl\'ement, A.~Blumen, J.~Rajchenbach, and J.~Duran.
\newblock Anomalous energy dissipation in molecular dynamics simulations of
  grains: The ``detachment effect''.
\newblock {\em Phys. Rev. E}, 50:4113, 1994.

\bibitem{zhang_jchp_13}
K.~Zhang, M.~Wang, S.~Papanikolaou, Y.~Liu, J.~Schroers, M.~D. Shattuck, and
  C.~S. O'Hern.
\newblock Computational studies of the glass-forming ability of model bulk
  metallic glasses.
\newblock {\em J. Chem. Phys.}, 139(12):124503, 2013.

\bibitem{Miller2004}
S.~Miller and S.~Luding.
\newblock Cluster growth in two- and three-dimensional granular gases.
\newblock {\em Phys. Rev. E}, 69(3 1):031305--1--031305--8, 2004.

\bibitem{Poeschel2005}
T.~P\"oschel, N.V. Brilliantov, and T.~Schwager.
\newblock Transient clusters in granular gases.
\newblock {\em J. Phys. Condens. Matter}, 17:S2705--S2713, 2005.

\bibitem{dapeng}
D.~Bi, J.~Zhang, B.~Chakraborty, and R.~P. Behringer.
\newblock Jamming by shear.
\newblock {\em Nature}, 480:355–358, 2011.

\bibitem{Zhang_soft_2010}
J.~Zhang, T.~S. Majmudar, M.~Sperl, and R.~P. Behringer.
\newblock Jamming for a 2d granular material.
\newblock {\em Soft Matter}, 6:2982--2991, 2010.

\bibitem{sastry_nature16}
Vinutha~H. A. and S.~Luding.
\newblock Disentagling the role of structure and friction in shear jamming.
\newblock {\em Nature Physics}, 12:578 -- 583, 2016.

\bibitem{luding_nature16}
S.~Luding.
\newblock So much for the jamming point.
\newblock {\em Nature Physics}, 12:531--532, 2016.

\bibitem{Minh2014}
N.~H. Minh, Y.~P. Cheng, and C.~Thornton.
\newblock Strong force networks in granular mixtures.
\newblock {\em Granul. Matter}, 16:69--78, Feb 2014.

\bibitem{mueth98}
D.~M. Mueth, H.~M. Jaeger, and S.~R. Nagel.
\newblock Force distribution in a granular medium.
\newblock {\em Phys. Rev. E}, 57(3):3164--3169, 1998.

\bibitem{radjai_prl_96}
F.~Radjai, M.~Jean, J.-J. Moreau, and S.~Roux.
\newblock Force distributions in dense two-dimensional granular systems.
\newblock {\em Phys. Rev. Lett.}, 77:274--278, 1996.

\bibitem{azema_pre_07}
E.~Az\'ema, F.~Radja\"{\i}, R.~Peyroux, and G.~Saussine.
\newblock Force transmission in a packing of pentagonal particles.
\newblock {\em Phys. Rev. E}, 76:011301, 2007.

\bibitem{makse_prl_00}
H~A. Makse, D.~L. Johnson, and L.~M. Schwartz.
\newblock Packing of compressible granular materials.
\newblock {\em Phys. Rev. Lett.}, 84:4160--4163, 2000.

\bibitem{cohen_2004}
R.~Cohen and S.~Havlin.
\newblock Fractal dimensions of percolating networks.
\newblock {\em Physica A Stat. Mech. Appl.}, 336:6 -- 13, 2004.

\bibitem{Newman_2010}
M.~Newman.
\newblock {\em Networks: An Introduction}.
\newblock Oxford University Press, Inc., New York, NY, USA, 2010.

\bibitem{Liu}
A.J. Liu and S.R. Nagel.
\newblock The jamming transition and the marginally jammed solid.
\newblock {\em Ann. Rev. of Cond. Matt. Phys.}, 1:347--369, 2010.

\bibitem{hecke_2010}
M.~van Hecke.
\newblock Jamming of soft particles: geometry, mechanics, scaling and
  isostaticity.
\newblock {\em J. Phys. Condens. Matter}, 22(3):033101, 2010.

\bibitem{Henkes_epl_2010}
S.~Henkes, M.~van Hecke, and W.~van Saarloos.
\newblock Critical jamming of frictional grains in the generalized isostaticity
  picture.
\newblock {\em EPL (Europhys. Lett.)}, 90(1):14003, 2010.

\bibitem{majmudar07a}
T.~S. Majmudar, M.~Sperl, S.~Luding, and R.~P. Behringer.
\newblock The jamming transition in granular systems.
\newblock {\em Physical Review Letters}, 98:058001--058004, 2007.

\bibitem{Shundyak_pre_2007}
K.~Shundyak, M.~van Hecke, and W.~van Saarloos.
\newblock Force mobilization and generalized isostaticity in jammed packings of
  frictional grains.
\newblock {\em Phys. Rev. E}, 75:010301, 2007.

\bibitem{capstone}
L.~Kondic.
\newblock Capstone {L}aboratory {W}eb {P}age, 2018.

\end{thebibliography}

\end{document}